\documentclass[aps,prb,twocolumn,showpacs,superscriptaddress,floatfix,nofootinbib]{revtex4-1}
\usepackage{soul}
\usepackage{graphicx}
\usepackage{amsmath}
\usepackage{epsfig}
\usepackage{helvet}
\usepackage{amssymb}

\newcommand{\be}{\begin{equation}}
\newcommand{\ee}{\end{equation}}
\newcommand{\bea}{\begin{eqnarray}}
\newcommand{\eea}{\end{eqnarray}}
\newcommand{\nn}{\nonumber}
\newcommand{\Tr}{\mathop{\mathrm{Tr}}}

\newcommand{\bra}[1]{\mbox{$\langle #1 |$}}
\newcommand{\ket}[1]{\mbox{$| #1 \rangle$}}
\newcommand{\braket}[2]{\mbox{$\langle #1 | #2 \rangle$}}

\newcommand{\mysymbol}[1]{\mathord{\includegraphics[scale = 1.3]{#1}}}
\newcommand{\al}{\alpha}

\newcommand{\GS}{\mbox{\tiny GS}}

\begin{document}
\title{Faster convergence of imaginary time evolution tensor network algorithms \\
by recycling the environment}

\author{Ho N. Phien}  \affiliation{Centre for Health Technologies, Faculty of Engineering and Information Technology, University of Technology Sydney, Sydney 2007, Australia}
\affiliation{Centre for Engineered Quantum Systems, School of Mathematics and Physics, University of Queensland, Brisbane 4072, Australia}

\author{Ian P. McCulloch} \affiliation{Centre for Engineered Quantum Systems, School of Mathematics and Physics, University of Queensland, Brisbane 4072, Australia}

\author{Guifr\'e Vidal} \affiliation{Centre for Engineered Quantum Systems, School of Mathematics and Physics, University of Queensland, Brisbane 4072, Australia}\affiliation{Perimeter Institute for Theoretical Physics, Waterloo, Ontario, N2L 2Y5, Canada}

\date{\today}
\begin{abstract}
We propose an environment recycling scheme to speed up a class of tensor network algorithms that produce an approximation to the ground state of a local Hamiltonian by simulating an evolution in imaginary time. Specifically, we consider the time-evolving block decimation (TEBD) algorithm applied to infinite systems in 1D and 2D, where the ground state is encoded, respectively, in a matrix product state (MPS) and in a projected entangled-pair state (PEPS).  An important ingredient of the TEBD algorithm (and a main computational bottle-neck, especially with PEPS in 2D) is the computation of the so-called \emph{environment}, which is used to determine how to optimally truncate the bond indices of the tensor network so that their dimension is kept constant. In current algorithms, the environment is computed at each step of the imaginary time evolution, to account for the changes that the time evolution introduces in the many-body state represented by the tensor network. Our key insight is that close to convergence, most of the changes in the environment are due to a change in the choice of gauge in the bond indices of the tensor network, and not in the many-body state. Indeed, a consistent choice of gauge in the bond indices confirms that the environment is essentially the same over many time steps and can thus be re-used, leading to very substantial computational savings. We demonstrate the resulting approach in 1D and 2D by computing the ground state of the quantum Ising model in a transverse magnetic field. 
\end{abstract}

\maketitle

%

\section{Introduction\label{secI}}

The study of interacting quantum many-body systems remains a central problem in modern physics. Understanding how the microscopic degrees of freedom organize themselves collectively is key to explaining the large variety of complex, emergent phenomena exhibited by quantum many-body systems. The study of such systems, for instance on a lattice, faces a major computational challenge, given the exponential growth of the Hilbert space dimension with the lattice size. There are several numerical approaches available, including exact diagonalization of the Hamiltonian on small lattices or quantum Monte Carlo, based on sampling an imaginary time evolution in Hamiltonians that are free from the sign problem. Among these and other methods, tensor network approaches stand up for their ability to approximating the full ground-state wave function, which is efficiently stored in a tensor network state.

The matrix product states (MPS) \cite{Fannes1,Ostlund1} is the best known example of a tensor network state. It is the basis of the density matrix renormalization group (DMRG) method \cite{White1, White2} to efficiently represent the ground state of a 1D quantum system in both finite and infinite\cite{Takasaki1, schollwock1, Perez1, Ian1, Ian2, Greg, schollwock2} lattices. The MPS is also the basis of methods to simulate time evolution, including the time evolving block decimation (TEBD) algorithm \cite{Vidal1,Vidal2} and its DMRG-like formulation \cite{tdDMRGWhite,tdDMRGSchollwock} which have been recently applied to study dynamical properties of 1D quantum system \cite{Phien1, Phien2, Zauner}, as well as the more recent algorithms based on the time-dependent variational principle \cite{TDVP}. Here we will focus on the TEBD algorithm, originally proposed for finite systems and subsequently extended to address systems in the thermodynamic limit, where it is referred to as the \emph{infinite} TEBD algorithm (iTEBD) \cite{Vidal3, Roman1}. In the thermodynamic limit, the many-body state is represented by an \emph{infinite} MPS (iMPS), which consists of a finite unit cell of MPS tensors that is repeated throughout the infinite chain. On the other hand, the projected entangled-pair state (PEPS)\cite{Verstraete1, Murg1, Verstraete2,Lubasch1,Lubasch2} or tensor product state (TPS) \cite{Nishino2, Nishino3, Gendiar1} was introduced as a natural generalization of the MPS for 2D systems.  In the thermodynamic limit, one can again use an infinite PEPS (iPEPS)\cite{Jordan1,Roman2}, which has a finite unit cell of tensors repeated throughout the infinite 2D lattice.

In both 1D and 2D, there are two main strategies to find a good approximation to the ground state of a local Hamiltonian $H$: one can either minimize the expectation value of $H$, succinctly referred to as \textit{energy minimization}, or one can simulate the action, on some initial state, of the evolution operator $e^{-tH}$ in imaginary time in the long time limit, referred to as \textit{imaginary time evolution}. In a 1D lattice, energy minimization (e.g., through the DMRG algorithm) is preferred over imaginary time evolution, because it generally converges faster to the ground state. In contrast, in 2D systems most current implementations are based on imaginary time evolution using the TEBD algorithm supplemented with a large variety of algorithms to contract the tensor network, including schemes based on a boundary MPS \cite{Jordan1}, the \textit{corner transfer matrix} \cite{Baxter11,Nishino1,Nishino1_1,Roman2,Philippe4,RomanNewCTM1}, and the \textit{tensor renormalization group} and its various important generalizations \cite{Levin1,Xie1,Zhao1,Xie2}. Among the most exciting explorations of ground states in challenging 2D systems by means of imaginary time evolution on iPEPS one can find e.g., recent calculations of the \emph{t-J}  model on the square lattice \cite{Philippe3,Philippe5} and honeycomb lattice \cite{Gu}, and of the $J_1-J_2$ frustrated Heisenberg model on the square lattice \cite{Wang}.
 
In order to simulate the action of the imaginary time evolution operator $e^{-tH}$ on some initial state, the TEBD algorithm proceeds by breaking this evolution operator into the product of a large number of operators $e^{-\delta H}$, each representing a small time step, $\delta \ll 1$, and by then applying each of these operator $e^{-\delta H}$ successively. Each application of a small time step $e^{-\delta H}$ results in the growth of the bond dimension of the tensors in the tensor network, which in turn results in a significant increase in computational costs. In order to keep the computational cost in check, it is thus important to truncate the bond index back to its original size. An optimal truncation of the bond index of a tensor (that is, a truncation that minimizes the error introduced in the many-body state), requires the computation of the so-called \textit{environment}, which contains information on how the truncation affects the rest of the many-body state. 

In a 1D system with open boundary conditions (OBC), the cost of computing the environment is similar to the cost of applying the time evolution $e^{-\delta H}$. However, in 2D, the cost of computing the environment is much larger than the cost of applying $e^{-\delta H}$, and in fact computing the environment is the major bottle-neck of the whole TEBD algorithm. So much so, that it has become a standard practice to bypass the computation of the environment altogether, and use instead just local information as a guide to perform the truncation of the bond indices, in what is known as the \textit{simplified update} (SU)\cite{JiangSimplify1, LiSimplify1}. In spite of its wide use within the PEPS community, the SU is not well-justified, and it is known to sometimes lead to poor PEPS approximations of the ground state. In other words, the SU, with its lower computational cost, allows to use a PEPS with larger bond dimension; however it does not exploit this larger bond dimension in an optimal way, potentially producing worse approximations to the ground state than what one would have obtained with a properly optimized PEPS with smaller bond dimension. Therefore the so-called \textit{full update} (FU), which requires the computation of the environment, is still the most reliable option in spite of its prohibitive cost.

In this paper, we thus focus on simulating imaginary time evolution using TEBD with FU, and propose a strategy to reduce its computational cost. Our strategy aims mainly at speeding up PEPS computations of 2D ground states, but it is also be useful in MPS calculations in 1D systems with periodic boundary conditions (PBC) as opposed to the OBC case alluded to above, where computing the environment is also significantly more expensive than just applying $e^{-\delta H}$. The key observation is that while at the beginning of the imaginary time evolution the many-body state is changing rapidly, in later iterations of the TEBD algorithm the changes are much less significant. This may not be directly apparent in the environment, which might still be subject to big changes. However, these big changes correspond mostly to a random choice in the gauge freedom existing in the tensor network. Specifically, on each bond index connecting two tensors of the MPS or PEPS, one always introduce the product of an invertible matrix and its inverse without affecting the represented many-body state. Moreover, it is important to remember that the environment is only used to determine the right choice of truncation of the bond dimension, and that a small change in the environment will not change this choice significantly. Therefore, if we manage to fix the gauge freedom in the environment so that an equivalent gauge is chosen over several time steps, then we may be able to compute the environment only \textit{once} and then recycle or re-use it to guide the truncation during \textit{many} time steps. As we will see, this strategy actually leads to very significant computational savings.

In order to be able to recycle the environment, we first need to bring the tensor network into a canonical form that fixes the gauge degrees of freedom in the bond indices. For an MPS with OBC, there is already a natural canonical form as well as a well-understood procedure to obtain it. For an MPS with PBC and a PEPS there is no obvious choice of a canonical form and it is not \textit{a priori} clear that a canonical form even exists. Here we propose an iterative procedure that, when applied to an MPS with OBC, has the usual canonical form as its fixed point. The same iterative procedure can then be applied to an MPS with PBC and (after suitable generalization) to a PEPS, and we verify numerically that a unique fixed-point is reached in each case. We then define the canonical form of an MPS with PBC and of a PEPS to be this fixed-point. For the purpose of incorporating the environment recycling schemes into the TEBD algorithm, we introduce an iterative method to obtain the canonical form of an MPS and a PEPS\cite{NoteAdded}. We notice that this type of canonical form of an PEPS is similar to the ``quasi-canonical'' form used in Refs.~\onlinecite{Kalis1,RomanNewCanonical1} to study the ground state of a 2D system by means of the SU.

For concreteness, in this work we discuss the recycling of the environment in a ground state computation by imaginary time evolution in an infinite system both in 1D and 2D, using iMPS and iPEPS, respectively. However, we emphasize that environment recycling can also be used in a more broader class of algorithms. On the one hand, one can recycle the environment when simulating imaginary time evolution on finite systems, even though such algorithms are not currently very popular. On the other hand, the very same ideas can also be applied to the computation of a classical partition function in 2D and 3D, using a scheme where a tensor network (e.g., an iMPS and an iPEPS, respectively) is used to encode the dominant eigenvector of a (1D or 2D, respectively) transfer matrix by a power method, where now this transfer matrix replaces the imaginary time evolution operator $e^{-\delta H}$. Indeed, after applying this transfer matrix a few times, the tensor network starts to approach the dominant eigenvector; from that moment on, the corresponding environment will no longer change significantly over iterations and it can be recycled.

Let us next summarize the structure of the paper. In Sec.~\ref{secII}, we briefly introduce MPS and PEPS. In Sec.~\ref{secIII}, we revise the canonical form of an iMPS and introduce an alternative iterative method to obtain it. We also provide methods to fix the remaining gauge degrees of freedom (mostly, complex phases) within a canonical iMPS. Then in Sec.~\ref{secIV} we propose a canonical form for iPEPS and a scheme to fix all its gauge degrees of freedom. In Sec.~\ref{secV} we present an environment recycling scheme for TEBD on iMPS and some benchmark results for 1D quantum Ising model with transverse magnetic field. Next, in Sec.~\ref{secVI} the environment recycling scheme is developed for TEBD on iPEPS and applied to study 2D quantum Ising model with the transverse magnetic field. Finally, Sec.~\ref{secVII} contains our conclusions.

{\it Note added}: In this paper, we use the terms MPS and PEPS to denote tensor networks in 1D and 2D, respectively, and the term TEBD to denote an algorithm that simulates a time evolution (in imaginary time) according to $e^{-tH}$ in order to obtain an MPS/PEPS approximation to the ground state of $H$. In the literature one often finds that the term \textquotedblleft iPEPS algorithm\textquotedblright is used to denote what we here call instead the TEBD algorithm for iPEPS. In this paper, we find it important to distinguish between the tensor network being used (e.g., MPS, PEPS, or even iMPS, iPEPS) and the algorithm that is employed to optimize it (DMRG, TEBD, etc.).

\section{Tensor network states \label{secII}}

In this section, we briefly revise the matrix product state and the projected entangled-pair state employed to represent the many-body state of 1D and 2D quantum lattice systems respectively.
\subsection{Matrix product states}
Let us consider a 1D quantum lattice system consisting of $N$ sites. Each lattice site has an internal degree of freedom, for example, a spin, and is represented by a local $d$-dimensional Hilbert space $\mathcal{H}_{d}=\mathbb{C}^{d}$. The $N$-site lattice is then described in the Hilbert space $\mathcal{H} = (\mathcal{H}_{d})^{\otimes N}$. A general pure state of the system can be written in the local basis as 
\bea
\ket{\Psi}=\sum_{s_{1}s_{2}\ldots s_{N}=1}^{d}c_{s_{1}s_{2}\ldots s_{N}}\ket{s_{1},s_{2},\ldots,s_{N}},
\label{eqC3_1}
\eea
where $\ket{s_{1},s_{2},\ldots,s_{N}}=\ket{s_{1}}\otimes\ket{s_{2}}\otimes\ldots\otimes\ket{s_{N}}$, and $\ket{s_{i}}$ (for $i = 1,\ldots, N$ and $s_i = 1,\ldots, d$) is the orthonormal basis of the local $d$-dimensional Hilbert space at site $i$, and $c_{s_{1}s_{2}\ldots s_{N}}$ is a complex component of a rank-$N$ tensor $c$. The total number of components $c_{s_{1}s_{2}\ldots s_{N}}$ is $d^{N}$, and hence proliferates exponentially with the number of lattice sites. 

\begin{figure}[htpb]
\begin{center}
\includegraphics[scale = 1]{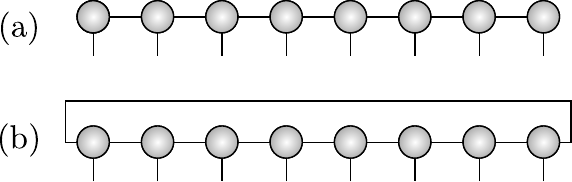}
\caption{ (Color online) Diagrammatic representations of an MPS corresponding to different boundary conditions: (a) The MPS with OBC. (b) The MPS with PBC. In these representations, a tensor is represented by a circle which has \emph{legs} corresponding to the indices of the tensor. The links connecting tensors together represent the contraction over bond indices. The open legs of the tensors correspond to the physical indices, which are not shared between tensors in the network.}
\label{fig1}
\end{center}\end{figure}

We can decompose the rank-$N$ tensor $c$ into a product of lower-rank tensors without changing the state of the system. More precisely,  if the lattice has OBC, the pure state in Eq.~(\ref{eqC3_1}) can be rewritten as follows:
\bea
\ket{\Psi} = \sum_{s_{1}s_{2}\ldots s_{N}=1}^{d}A^{s_{1}}A^{s_{2}}\ldots A^{s_{N}}\ket{s_{1},s_{2},\ldots,s_{N}},
\label{eqC3_2}
\eea
where $A^{s_{1}}$ and $A^{s_{N}}$ corresponding to each value of $s_{1}$ and $s_{N}$ are the $1\times\chi$ row and $\chi\times 1$ column vectors, respectively, and each $A^{s_{k}}$ (for $k=2,\ldots,N-1$) corresponding to each value of $s_{k}$ is a $\chi\times\chi$ matrix. The pure state represented by  Eq.~(\ref{eqC3_2}) is called as a \emph{matrix product state}; see Fig.~\ref{fig1}(a). This MPS is parameterized by $(N-2)\chi^{2}d +2\chi d$ variational parameters, and accordingly the state is represented by a number of parameters, which increases polynomially with $N$. The coeficient $\chi\geq 1$ is defined as the \emph{bond dimension} and understood as a  \emph{refinement} parameter. More precisely, $\chi$ plays an important role in qualifying the MPS representation as the larger the $\chi$, the better the MPS representation. If $\chi = 1$, the MPS represents a product state otherwise the MPS describes an entangled state. We can always choose an upper bound of $\chi$ to represent exactly the general state of $N$-site lattice, say $\chi = d^{N/2}$, but this will result in poor use of computational resources because the MPS is no longer an efficient representation.

In case that the lattice obeys PBC, the MPS expressed by Eq.~(\ref{eqC3_2}) can still represent correctly the state of the system. This is straightforward because the tensor $c$ decomposed into a product of tensors $\{A^{s_{k}}\}_{k=1}^{N}$ does not impose any specific boundary condition on the lattice. However, it is not a feasible representation in the sense that there is no connection between the tensors representing the first and the last sites of the lattice. Therefore the correlation between these two sites is not transmitted through the link of the system. Alternatively, we can generalize this representation where all the tensors including the first and the last ones have two bond indices. The generalized MPS is then written as follows,
\bea
\ket{\Psi} =\sum_{s_{1}s_{2}\ldots s_{N}=1}^{d} \Tr(A^{s_{1}}A^{s_{2}}\ldots A^{s_{N}})\ket{s_{1},s_{2},\ldots,s_{N}}.
\label{eqC3_3}
\eea
Note that the \emph{trace} (Tr) is taken to ensure that after contracting all the matrices $\{A^{s_{k}}\}_{k=1}^{N}$ we have a scalar quantity. A diagrammatic representation of the MPS with PBC is shown in Fig.~\ref{fig1}(b), where the trace is taken by connecting the left bond of tensor $A^{s_{1}}$ with the right bond of tensor $A^{s_{N}}$.

In reality, the representation of an MPS with PBC often causes some computational difficulties in simulation. More explicitly, a simulation using MPS with PBC is much more computationally expensive than employing MPS with OBC. As the cyclic structure of the MPS with PBC, it requires larger computational resources to capture the amount of entanglement in the system. The computational cost scales as $\mathcal{O}(\chi^{5})$ \cite{Verstraete1_2} comparing to $\mathcal{O}(\chi^{3})$ when simulating using MPS with OBC. However, this scaling sometimes can be reduced to $\mathcal{O}(\chi^{3})$ \cite{SandvikPBC1,PippanPBC}.

Note that MPS can also represent the state of an infinite-size homogeneous system using only a small number of parameters by employing its invariance under translations. This results in what is called \emph{infinite} MPS (iMPS). An iMPS is normally constructed by repeatedly assigning a unit cell made up of a finite set of tensors along the lattice chain. For example, a one-site iMPS is represented as follows:
\bea
\ket{\Psi}&=&\sum_{\ldots s\ldots=1}^{d}\ldots\lambda\Gamma^{s}\ldots\ket{\ldots, s,\ldots}.
\label{eqC3_4_18_iMPS}
\eea
In this iMPS, the unit cell contains a pair of tensors $\{\Gamma,\lambda\}$ where the tensor $\Gamma$ is positioned at each lattice site and the diagonal matrix $\lambda$ lies on the link between each two adjacent sites. The iMPS is respresented by only $\chi^2d +\chi$ variational parameters. This is the most striking advantage of using translational invariant iMPS because we only need a small number of parameters to represent the ground state of a lattice system in the thermodynamic limit. Consequently, using an iMPS to represent a state when simulating a 1D infinite system obviously reduces the computational complexity as well as the computer memory.
\subsection{Projected entangled-pair states}
\begin{figure}[htpb]
\begin{center}
\includegraphics[width = \columnwidth]{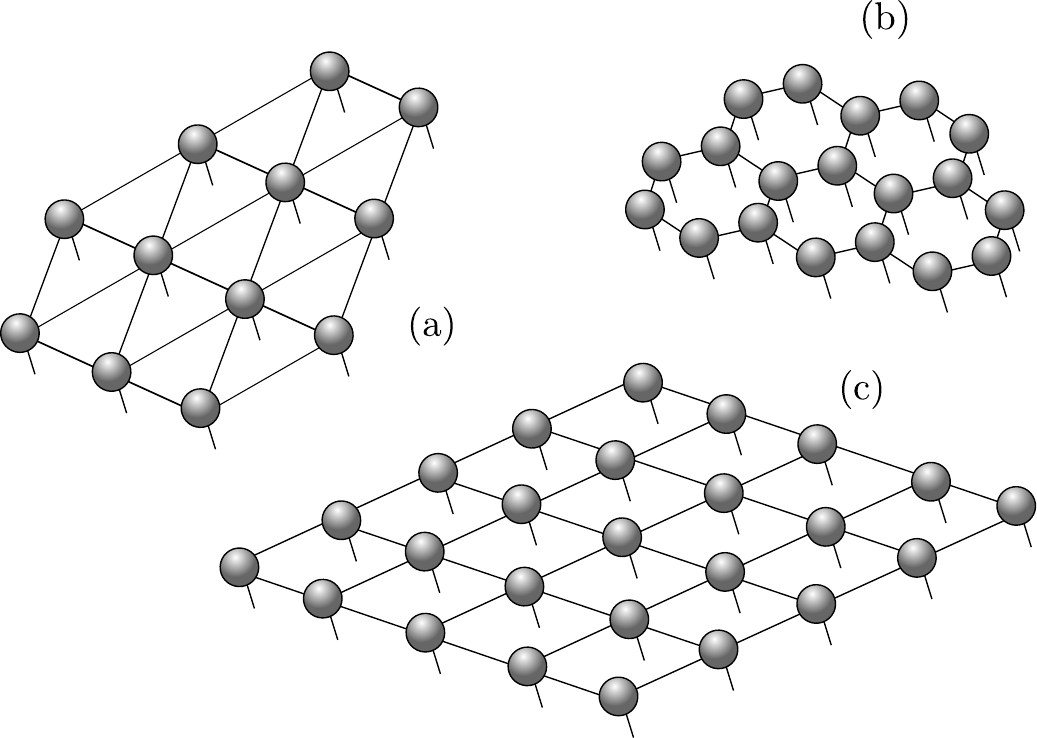}
\caption{ (Color online) Diagrammatic representations of PEPSs that correspond to different lattice patterns: (a) a triangular lattice, (b) a hexagonal lattice, and (c) a square lattice.}
\label{fig2}
\end{center}\end{figure}

Consider a 2D quantum system on a lattice of $N$ sites where each lattice site is represented by a local Hilbert space $\mathcal{H}_{d}\cong\mathbb{C}^{d}$. To represent the system in terms  of a TN, one option is to use the so-called \emph{projected entangled-pair state}\cite{Verstraete1,Murg1,Verstraete2} which is generally defined as
\bea
\ket{\Psi} = \sum_{s_{\vec{r}_{1}},\ldots, s_{\vec{r}_{N}}=1}^{d}F(A^{[\vec{r}_{1}]s_{\vec{r}_{1}}},\ldots, A^{[\vec{r}_{N}]s_{\vec{r}_{N}}})\ket{s_{\vec{r}_{1}},\ldots,s_{\vec{r}_{N}}}.\nn\\
\label{eqC4_1}
\eea
This PEPS is represented by tensors $\{A^{[\vec{r}_{i}]s_{\vec{r}_{i}}}\}_{i=1}^{N}$ (where $\vec{r}_{i} = (x_{i},y_{i})$ is the coordinate of the lattice site $i$), which are connected to some set of neighbor sites according to the geometry of the lattice. Each tensor of the PEPS has $n$ (the coordination number, or the number of nearest neighbors of one lattice site) bond indices of dimension $D$ and one physical index of dimension $d$. The choice of $n$ in the TN depends on the geometry of the lattice, and can in principle be chosen arbitrarily (although this affects the computational complexity of contracting the TN). The function $F$ contracts all the tensors  $\{A^{[\vec{r}_{i}]s_{\vec{r}_{i}}}\}_{i=1}^{N}$ according to this pattern and then performs the trace to obtain a scalar quantity. In Fig.~\ref{fig2}, we show diagrammatically several PEPSs for the systems corresponding to different geometries with OBC. In case the geometry of a system is a square lattice pattern with PBC ($n=4$), the PEPS consists of tensors that have four bond indices and one physical index. Overall, the PEPS depends on $\mathcal{O}(ND^4d)$ variational parameters. Note that the bond dimension $D$ determines the quality of PEPS, and if $D$ is chosen to be large enough, any state of a 2D lattice system can be well-represented by a PEPS.

As a generalization of the MPS representation to 2D, PEPS is a good representation for a class of variational wave functions when studying the ground state of 2D quantum lattice systems. The rationale behind this is that a PEPS is constructed naturally to capture the geometry of the lattice pattern, such that the scaling of entanglement entropy of a subregion is proportional to the surface area of the subregion, it fulfills the \emph{area law}\cite{Bekenstein1, Srednicki1, Eisert1}. More precisely, the entanglement entropy of an $l\times l$ block scales as the number of bonds that are cut in the TN, which is $\mathcal{O}(l)$ \cite{Verstraete3}. Thus a PEPS can represent efficiently the ground state of a 2D quantum lattice system with a small number of parameters which scales at worst polynomially with the system size. Furthermore, it can also effectively capture the quantum correlations existing in the system. There is some prospect of capturing also dynamical properties, as long as the dynamical growth in entanglement entropy is not too large. Besides, as similar to MPS, we can also employ an \emph{infinite} PEPS (iPEPS) to represent a state of a 2D homogeneous system, which is parameterized by a fairly small number of parameters. 

\section{Canonical form of an iMPS and gauge degrees of freedom fixing\label{secIII}}

In this section we revise the canonical form of an iMPS and introduce an iterative method to obtain it. This method as we will see can be generalized to apply to MPS with either OBC or PBC. Besides, some methods applied to fix the gauge degrees of freedom in a canonical iMPS will also be introduced explicitly.

\subsection{Canonical form of an iMPS}
For simplicity, let us consider an infinite lattice chain represented by a one-site translationally invariant iMPS in Eq.~(\ref{eqC3_4_18_iMPS}). This iMPS is said to be in its canonical form if $\lambda$ contains the Schmidt coefficients and therefore we can always write it in the Schmidt decomposition at any bond of the chain as follows,
\bea
\ket{\Psi} = \sum_{\al=1}^{\chi}\lambda_{\al}\ket{\phi^{\al}}_{L}\ket{\phi^{\al}}_{R},
\label{SchmidtC7}
\eea
with $\ket{\phi^{\al}}_{L}$ and $\ket{\phi^{\al}}_{R}$ being orthonormal bases. Equivalently, this iMPS must satisfy the orthonormality constraints defined as,
\bea
\label{iMPS_canonical_constraintL}
\sum_{s}{\Gamma^{s}}^{\dagger}\rho_{L}\Gamma^{s}=\mathbb{I},\\
\sum_{s}\Gamma^{s}\rho_{R}{\Gamma^{s}}^{\dagger}=\mathbb{I},
\label{iMPS_canonical_constraintR}
\eea
where $\rho_{L}=\rho_{R} =\lambda^{2}$ and $\mathbb{I}$ are reduced density and identity matrices respectively.

\begin{figure}[htpb]
\begin{center}
\includegraphics[width=\columnwidth]{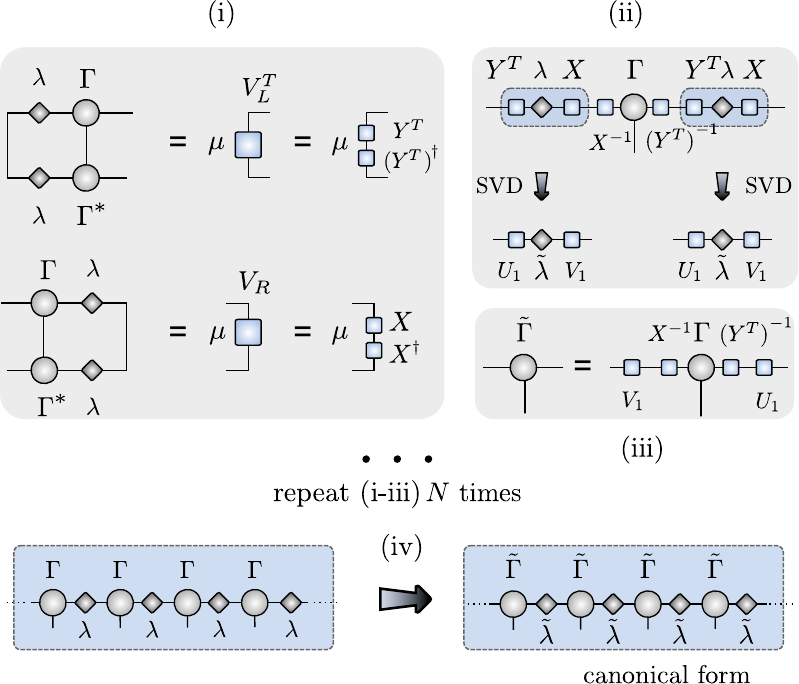}
\caption[(i-iii) Fundamental steps applied to transfer an iMPS $\{\Gamma,\lambda\}$ into its canonical form]{(i-iii) Fundamental steps applied to transfer an iMPS $\{\Gamma,\lambda\}$ into its canonical form. These steps are iterated until the Schmidt coefficients converge. (iv) Two iMPSs $\{\Gamma,\lambda\}$ and $\{\tilde{\Gamma},\tilde{\lambda}\}$ are different from each other but they represent the same physical state $\ket{\Psi}$.}
\label{fig3}
\end{center}\end{figure}

To transfer an arbitrary iMPS represented by  Eq.~(\ref{eqC3_4_18_iMPS}) into a canonical form, one can apply the scheme introduced in  Ref.~[\onlinecite{Roman1}]. Here, as an alternative, we propose an iterative method that can also handle well this task. The key idea of this iterative method is quite similar to the one in Ref.~[\onlinecite{Roman1}]. The main difference is that we do not need the left- and right-dominant eigenvectors of a transfer matrix to find the resolutions but instead we will find the resolutions locally. More concretely, our scheme consists of the following fundamental steps:
\begin{itemize}
 \item[(i)] Compute the matrices $V_{L}$ and $V_{R}$ whose coefficients are defined as follows:
\bea
V_{L} &=& \sum_{s}{(\lambda\Gamma^{s})^{\dagger}(\lambda\Gamma^{s})},\\
V_{R} &=& \sum_{s}{(\Gamma^{s}\lambda)(\Gamma^{s}\lambda)^{\dagger}},
\eea
which are shown graphically in Fig.~\ref{fig3}(i). As $V_{L}$ and $V_{R}$ are Hermitian and positive matrices, we can decompose $V_{L} = Y^{\dagger}Y$ and $V_{R} = XX^{\dagger}$ by means of the eigenvalue decomposition. For instance, we can decompose $V_{L} = WSW^{\dagger}$ and then assign $Y^{\dagger} = W\sqrt{S}$ and $Y = \sqrt{S}W^{\dagger}$. Similarly, we can easily obtain $X = W'\sqrt{S'}$ and $X^{\dagger} = \sqrt{S'}W'^{\dagger}$ from the decomposition $V_{R} = W'S'W'^{\dagger}$.
\item[(ii)] Insert two resolutions $\mathbb{I} = (Y^{T})^{-1}Y^{T}$ and $\mathbb{I} = XX^{-1}$ into the bonds of the iMPS as illustrated in Fig.~\ref{fig3}(ii). Then we get a new $\tilde{\lambda}$ which contain the singular values from taking the SVD of $Y^{T}\lambda X$ such that $U_{1}\tilde{\lambda}V_{1} = Y^{T}\lambda X$.
\item[(iii)] A new tensor $\tilde{\Gamma}$ is obtained by contracting all the tensors including $V_{1}, X^{-1}, \Gamma, (Y^{T})^{-1}$ and $U_{1}$ together; see Fig.~\ref{fig3}(iii). Re-assign $\Gamma\equiv\tilde{\Gamma}$ and $\lambda\equiv\tilde{\lambda}$ then go back to step (i). 
\end{itemize}
The above steps are repeated until the spectrum consisting of singular values $\tilde{\lambda}$ converge to a fixed point. Then the iMPS represented by $\{\tilde{\Gamma},\tilde{\lambda}\}$ fulfills the conditions in Eqs.~(\ref{iMPS_canonical_constraintL},\ref{iMPS_canonical_constraintR}), and thus is in the canonical form.

We have empirically observed that the above algorithm always converged to the canonical form. Note also that because the canonical form of an iMPS is obtained locally without taking into account the information of the whole system, we can also apply this scheme to find the canonical form for MPS with PBC when it is translationally invariant under shifts of a certain unit cell of sites. The canonical form of the MPS with PBC seems to be closely related to the one defined in Ref.~\onlinecite{Perez1} for the case of translational invariant state. However, we are unable to confirm that they are exactly the same since in our case it can be also defined for the non-translational invariant state as well. 

\subsection{Fixing gauge degrees of freedom in an iMPS\label{subsecIIIB}}
We now introduce two possible methods to fix the gauge degrees of freedom between two different canonical iMPSs that represent the same physical state $\ket{\Psi}$. Suppose that the canonical iMPSs are one-site translationally invariant and denoted as $\ket{\Psi_{A}}$ and $\ket{\Psi_{B}}$. In this notation, $\ket{\Psi_{A}}$ and $\ket{\Psi_{B}}$ are characterized by single tensors $A$ and $B$, respectively. For simplicity, let us assume that $\ket{\Psi_{A}}$ and $\ket{\Psi_{B}}$ are normalized. Due to the canonical form representations in these iMPSs, we come up with the relation $A = U^{\dagger}BU$, where $U$ can be a unitary matrix or a diagonal matrix consisting of complex phases. To fix the gauge degrees of freedom between $\ket{\Psi_{A}}$ and $\ket{\Psi_{B}}$, we need to search for the matrix $U$. In the following, with the assumption that $U$ is unitary, we will explain in details how to obtain $U$ with two different methods.

\subsubsection{Direct method} 
Suppose that $X$ is the left-dominant eigenvector of the transfer matrix created by two tensors $A$ and $B$ with corresponding left-dominant eigenvalue $\mu$. We have the following equation:
\bea
\sum_{s = 1}^{d}\sum_{\al',\al = 1}^{\chi}{\big(B^{s}_{\beta'\al'}\big)}^{*}X_{\al'\al}A^{s}_{\al\beta} = \mu X_{\beta'\beta}.
\label{iMPSDM1}
\eea
Note that for a clear demonstration, we will incorporate graphical representations for some expressions in the text. For example, Eq.~(\ref{iMPSDM1}) is shown graphically as
\bea
\mysymbol{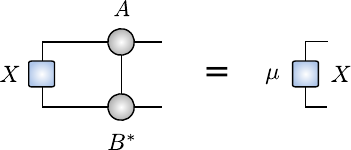}.\nn
\eea
Let us replace $A^{s}_{\al\beta} = \sum_{\gamma,\delta = 1}^{\chi}U^{*}_{\al\gamma}B^{s}_{\gamma\delta}U_{\delta\beta}$ into the above equation we have
\bea
\mysymbol{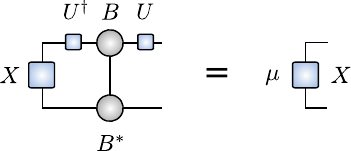}.
\label{eq121}
\eea
Multiplying both sides of Eq.~(\ref{eq121}) with $U^{\dagger}$ yields
\bea
\mysymbol{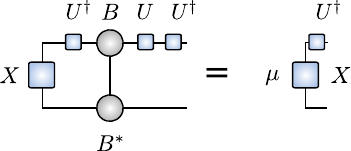},
\eea
which is simplified as
\bea
\mysymbol{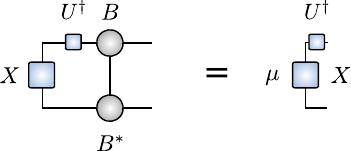}.
\label{gaugefreedomfixeq4}
\eea
We realize that $XU^{\dagger}$ is also an eigenvector of the transfer matrix created by tensors $B$ and $B^{\dagger}$ with corresponding eigenvalue $\mu$. However, as $\ket{\Psi_{B}}$ is defined in its canonical form, a possible solution of Eq.~(\ref{gaugefreedomfixeq4}) is $XU^{\dagger} = \mathbb{I}$. Thus, we can easily extract $U = X$ and $\mu = 1$. The matrix $U$ needed for fixing the gauge degrees of freedom between the two iMPSs $\ket{\Psi_{A}}$ and $\ket{\Psi_{B}}$ has been found exactly and this corresponds to the dominant eigenvector of the transfer matrix created by tensors $A$ and $B$.
\subsubsection{Iterative method \label{IterativeMethodiMPS}} 
We can also determine the matrix $U$ iteratively. To do that, let us define a \emph{local fidelity} as follows,
\bea
F&=&\sum_{s = 1}^{d}\sum_{\al,\beta,\gamma,\al' = 1}^{\chi}{U_{\al\beta}A^{s}_{\beta\gamma}(U_{\gamma\al'})^{*}(B^{s}_{\al'\al})^{*}},\nn\\
&=&\Tr(UAU^{\dagger}B^{\dagger})\nn\\
&=&\text{Tr}\big(UM\big).
\label{eq15}
\eea
This equation is visualized as
\bea
\mysymbol{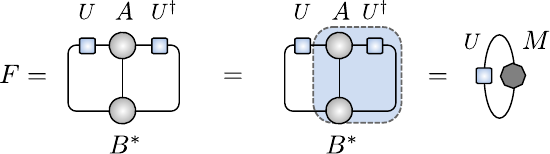}\nn,
\eea
where we have defined $M = AU^{\dagger}B^{\dagger}$. Our task is to find the matrix $U$ that maximizes the fidelity $F$. More specifically, we first initialize an arbitrary $U$, e.g., $U=\mathbb{I}$ and then iterate the following steps:
\begin{itemize}
 \item[(i)] Apply the SVD to decompose the matrix $M$ such that $M = WSV^{\dagger}$.
 \item[(ii)] Assign $U = VW^{\dagger}$ and substitute it back into Eq.~(\ref{eq15}).
 \item[(iii)] Compute the fidelity which is now defined as $F=\text{Tr}(S)$.
 \item[(iv)] Re-compute the matrix $M = AU^{\dagger}B^{\dagger}$ and then go back to step (i).
\end{itemize}
The above steps are iterated until the fidelity converges to a fixed point, i.e., $F=1$.

\section{Canonical form of an iPEPS and gauge degrees of freedom fixing\label{secIV}}
Similar to an iMPS, an iPEPS representing a pure state $\ket{\Psi}$ of a 2D lattice system is not unique, and there exists a lot of different iPEPSs representing the same state physical $\ket{\Psi}$. These iPEPSs can be distinguished by some gauge degrees of freedom. For instance, for a one-site translationally invariant iPEPS $\ket{\Psi_{A}}$, characterized by tensor $A$, one can always insert $M^{-1}M=\mathbb{I}$ and $K^{-1}K=\mathbb{I}$ ($M$ and $K$ are arbitrary invertible matrices) into \emph{horizontal} and \emph{vertical} links of the iPEPS $\ket{\Psi_{A}}$, respectively, without changing its physical properties. More concretely, we define a new iPEPS $\ket{\Psi_{B}}$ characterized by tensor $B$ given by
\bea
B^{s}_{lrdu} = \sum_{l'r'd'u'}M_{ll'}K_{dd'}A^{s}_{l'r'd'u'}M^{-1}_{r'r}K^{-1}_{u'u},
\eea
which describes a same physical state as $\ket{\Psi_{A}}$. The existence of gauge degrees of freedom sometimes causes lots of numerical difficulties in dealing with iPEPS representation. Therefore, we need to fix the gauge degrees of freedom in every iPEPS. This can be done by defining a new type of \emph{canonical form} for an iPEPS in what follows.
\subsection{Canonical form of an iPEPS}
\begin{figure}[htpb]
\begin{center}
\includegraphics[scale = 0.8]{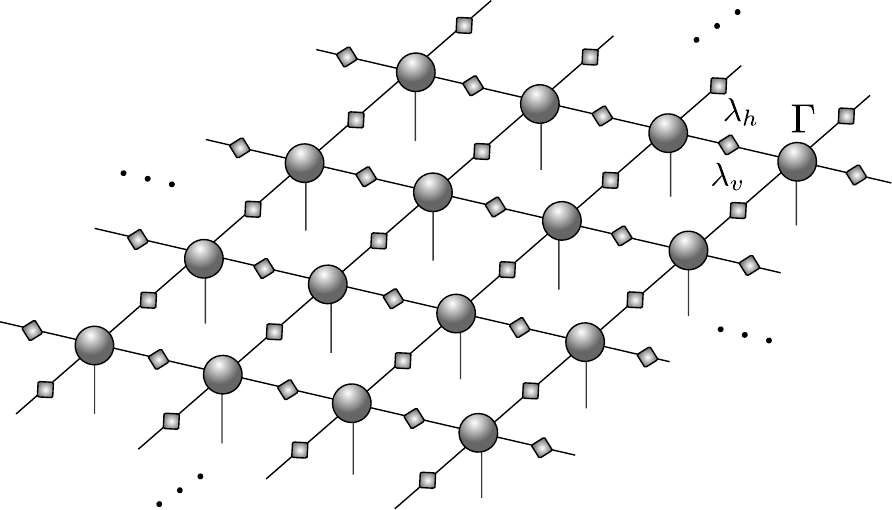}
\caption{ (Color online) Diagrammatic representation of an infinite one-site translationally invariant iPEPS which consists of tensors $\{\lambda_{h}, \lambda_{v}, \Gamma\}$.}
\label{fig9}
\end{center}\end{figure}

For the sake of simplicity, we consider a one-site translationally invariant iPEPS, which is represented by tensors $\{\Gamma,\lambda_{h}, \lambda_{v}\}$ and is visualised as in Fig.~\ref{fig9}. We can also represent the iPEPS with a single tensor $A$ by contracting the matrices  $\lambda_{h}$ and $\lambda_{v}$ to tensor $\Gamma$ such that $A = \lambda_{h}\lambda_{v}\Gamma$. 

Note that for a TN with closed loops such as an iPEPS, it is impossible to define a canonical form based on the orthonormality of a bipartite Schmidt decomposition. However, in what follows, we define a new type of canonical form for an iPEPS in the sense that the tensors in the unit cell of the iPEPS satisfy some particular constraints. This definition for a canonical form iPEPS is completely different from the one given in Ref.~\onlinecite{PerezCanonicalPEPS} where the canonical form of a \emph{finite} PEPS is termed from characterizing the existence of symmetries given by the fact that two representations of the same \emph{injective} PEPS are related by unique invertible matrices. In what follows, we define the canonical form based on the orthogonality constraints of local tensors of the iPEPS.

\begin{figure}[htpb]
\begin{center}
\includegraphics[width = \columnwidth]{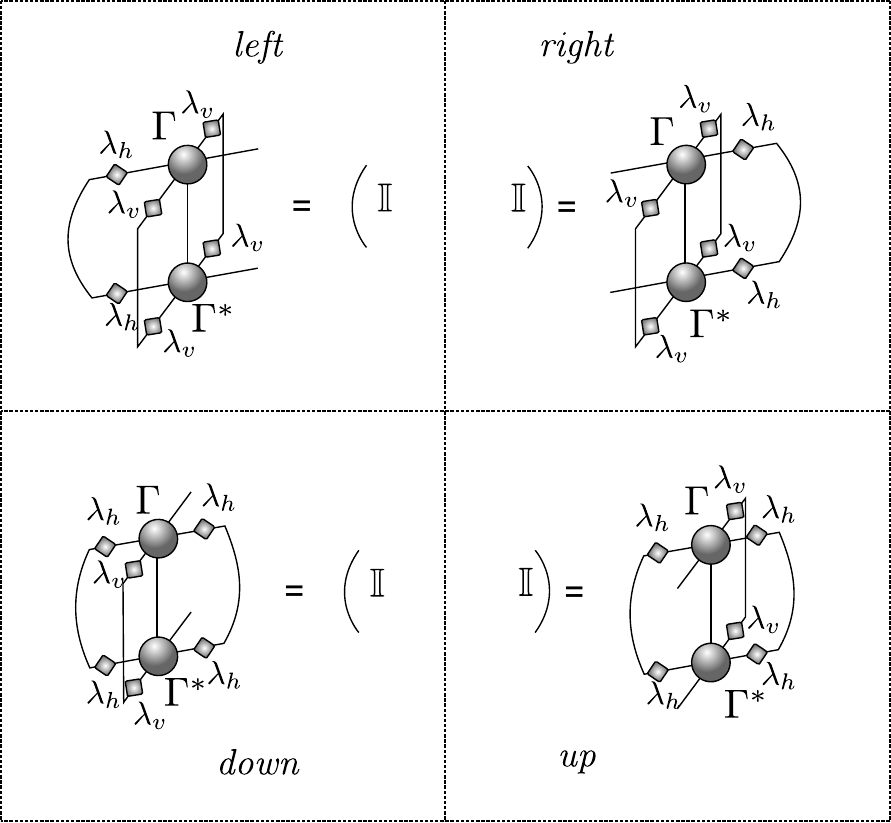}
\caption{ (Color online) Diagrammatic representation of constraints for the canonical iPEPS.}
\label{fig10}
\end{center}\end{figure}

\emph{Definition.} An iPEPS is defined in its canonical form if it is represented by the tensors $\{\Gamma,\lambda_{h}, \lambda_{v}\}$ satisfying the following constraints:
\bea
\label{eq17}
\sum_{s,l,d,u}L^{s}_{lrdu}({L^{s}}_{lr'du})^{*} &=& \delta_{rr'},\\
\label{eq18}
\sum_{s,r,d,u}R^{s}_{lrdu}({R^{s}}_{l'rdu})^{*} &=& \delta_{ll'},\\
\label{eq19}
\sum_{s,l,r,d}\mathcal{D}^{s}_{lrdu}({\mathcal{D}^{s}}_{lrdu'})^{*} &=& \delta_{uu'},\\
\label{eq20}
\sum_{s,l,r,u}U^{s}_{lrdu}({U^{s}}_{lrd'u})^{*} &=& \delta_{dd'},
\eea 
where we have  defined
\bea
L^{s}_{lrdu} &=& \sum_{l',d',u'}\Gamma^{s}_{l'rd'u'}(\lambda_{h})_{ll'}(\lambda_{v})_{dd'}(\lambda_{v})_{u'u},\\
R^{s}_{lrdu} &=& \sum_{r',d',u'}\Gamma^{s}_{lr'd'u'}(\lambda_{h})_{r'r}(\lambda_{v})_{dd'}(\lambda_{v})_{u'u},\\
\mathcal{D}^{s}_{lrdu} &=& \sum_{l',r',d'}\Gamma^{s}_{l'r'd'u}(\lambda_{h})_{ll'}(\lambda_{v})_{dd'}(\lambda_{h})_{r'r},\\
U^{s}_{lrdu} &=& \sum_{l',r',u'}\Gamma^{s}_{l'r'du'}(\lambda_{h})_{ll'}(\lambda_{v})_{u'u}(\lambda_{h})_{r'r}.
\eea
The conditions defined by Eqs.~(\ref{eq17})-(\ref{eq20}) can be named as the \emph{left, right, down,} and \emph{up} canonical constraints for the canonical form of the iPEPS and are shown diagrammatically in Fig.~\ref{fig10}.

\begin{figure}[htpb]
\begin{center}
\includegraphics[scale = 0.9]{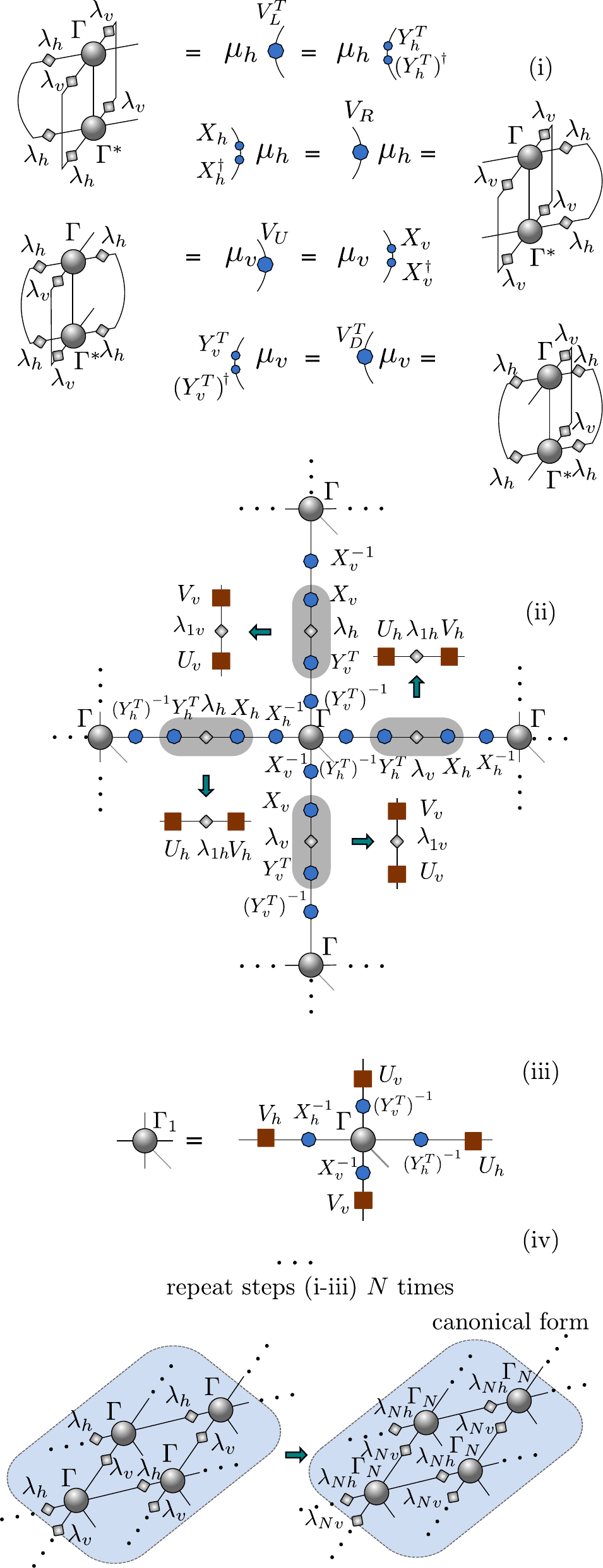}
\caption{ (Color online) Diagrammatical illustration of how to transfer an iPEPS into its canonical form.}
\label{fig11}
\end{center}\end{figure}

By generalizing the iterative method applied to transfer an iMPS into its canonical form, we can obtain the canonical form for an arbitrary iPEPS. In particular, we iterate the following steps which are also illustrated diagrammatically in Fig.~\ref{fig11},
\begin{itemize}
 \item[(i)] Compute matrices $V_{L}$, $V_{R}$, $V_{D}$ and $V_{U}$ to find the necessary resolutions; see Fig.~\ref{fig11}(i). As all of these matrices are Hermitian and positive matrices, we can decompose $V_{L} = Y_{h}^{\dagger}Y_{h}$, $V_{R} = X_{h}X_{h}^{\dagger}$, $V_{D} = Y_{v}^{\dagger}Y_{v}$, $V_{U} = X_{v}X_{v}^{\dagger}$ by employing, for example, the eigenvalue decomposition. Specifically, we can construct $V_{L} = W_{h}D_{h}W_{h}^{\dagger}$, $V_{R} = Q_{h}T_{h}Q_{h}^{\dagger}$, $V_{D} = W_{v}D_{v}W_{v}^{\dagger}$, and $V_{U} = Q_{v}T_{v}Q_{v}^{\dagger}$ then assign $Y_{h} = \sqrt{D_{h}}W_{h}^{\dagger}$, $X_{h} = Q_{h}\sqrt{T_{h}}$, $Y_{v} = \sqrt{D_{v}}W_{v}^{\dagger}$ and $X_{v} = Q_{v}\sqrt{T_{v}}$, respectively.
\item[(ii)] Insert simultaneously four resolutions $(Y_{h}^{T})^{-1}Y_{h}^{T}$, $X_{h}X_{h}^{-1}$, $(Y_{v}^{T})^{-1}Y_{v}^{T}$, and $X_{v}X_{v}^{-1}$, which are all  identity matrices, into the bonds of the iPEPS as illustrated in Fig.~\ref{fig11}(ii). Two new diagonal matrices $\tilde{\lambda}_{1h}$ and $\tilde{\lambda}_{1v}$ are then obtained by taking the SVDs of $Y_{h}^{T}\lambda_{h} X_{h}$ and $Y_{v}^{T}\lambda_{h} X_{v}$, respectively.
\item[(iii)] Define a new tensor $\tilde{\Gamma}$ which is obtained by contracting all the remaining tensors $V_{h}$, $X_{h}^{-1}$, $V_{v}$, $X_{v}^{-1}$, $\Gamma$, $(Y_{h}^{T})^{-1}$, $U_{h}$, $(Y_{v}^{T})^{-1}$, and $U_{v}$ together; see Fig.~\ref{fig11}(iii). Re-assign $\Gamma\equiv\tilde{\Gamma}$, $\lambda_{h}\equiv\tilde{\lambda}_{1h}$ and $\lambda_{v}\equiv\tilde{\lambda}_{1v}$ then go back to step (i). 
\end{itemize}
The above steps are repeated until the fixed point of the spectrum described by diagonal tensors $\tilde{\lambda}_{h},\tilde{\lambda}_{v}$ is reached. Then the iPEPS represented by tensors $\{\tilde{\Gamma},\tilde{\lambda}_{h},\tilde{\lambda}_{v}\}$ is in its canonical form and satisfies the conditions defined in Eqs.~[\ref{eq17}-\ref{eq20}]. In practice, we empirically see that the iteration for obtaining the canonical form always converges.

\subsection{Fixing gauge degrees of freedom in an iPEPS\label{sec:ch5sec2sec5}}
One of the most significant benefits of working with the canonical form of an iPEPS is that we can fix its gauge degrees of freedom that can reduce many numerical difficulties. Even if an iPEPS is in the canonical form, there is still a gauge freedom to choose the complex phases of each index of the tensors. Here, by assuming that two different iPEPSs describing the same quantum state are in the canonical forms, we suggest an iterative method to fix the gauge degrees of freedom between them. In particular, let us call these two iPEPSs as $\ket{\Psi_{A}}$ and $\ket{\Psi_{B}}$, which are characterized by tensors $\{\Gamma_{A}, \lambda_{Ah}, \lambda_{Av}\}$ and $\{\Gamma_{B},\lambda_{Bh}, \lambda_{Bv}\}$, respectively. We define two new tensors $A = \lambda_{Ah}^{1/2}\lambda_{Av}^{1/2}\Gamma_{A}\lambda_{Ah}^{1/2}\lambda_{Av}^{1/2}$ and $B = \lambda_{Bh}^{1/2}\lambda_{Bv}^{1/2}\Gamma_{B}\lambda_{Bh}^{1/2}\lambda_{Bv}^{1/2}$, which satisfy the normalization $\Tr(A^{\dagger}A) = \Tr(B^{\dagger}B) = 1$. Correspondingly, these two tensors also characterize $\ket{\Psi_{A}}$ and $\ket{\Psi_{B}}$ and are different from each other by phases or unitary matrices. More precisely, they are related by $A = U_{h}^{\dagger}U_{v}^{\dagger}BU_{h}U_{v}$ where we assume that $U_{h}$ and $U_{v}$ are unitary matrices. Fixing the gauge degrees of freedom between $\ket{\Psi_{A}}$ and $\ket{\Psi_{B}}$ requires determining matrices $U_{h}$ and $U_{v}$. By generalizing the iterative method proposed in Sec.~\ref{IterativeMethodiMPS}, we can find $U_{h}$ and $U_{v}$ such that they maximize the local fidelity, defined as,
\bea
F&=&\Tr(U_{h}U_{v}AU_{h}^{\dagger}U_{v}^{\dagger}B^{\dagger}),
\label{eq25}
\eea
which is graphically represented as
\bea
\mysymbol{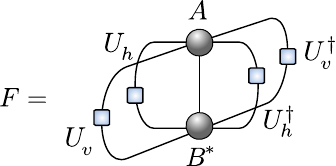}\nn.
\eea
More precisely, $U_{h}$ and $U_{v}$ are obtained iteratively as follows. To start, we initialize them arbitrarily, for example, we could start from identity matrices (or some random matrices), and repeat the following steps.
\begin{itemize}
 \item[(i)] Fix matrices $U_{v}$, $U_{h}^{\dagger}$, and $U_{v}^{\dagger}$ to search $U_{h}$ by computing the matrix $M_{h}$ such that
\bea
\mysymbol{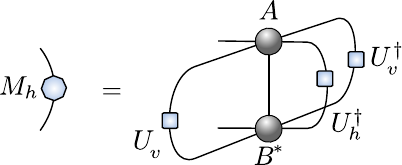}.
\eea 
 \item[(ii)] Find the matrix $U_{h}$ that might maximize the local fidelity (\ref{eq25}). Similar to the 1D case, $U_{h}$ can be determined via decomposing the matrix $M_{h}$ using SVD such that $M_{h}= W_{h}S_{h}V_{h}^{\dagger}$. We then set $U_{h} = V_{h}W_{h}^{\dagger}$.
 \item[(iii)] Update the new matrices $U_{h}$ and $U_{h}^{\dagger}$.
 \item[(iv)] Fix the matrices $U_{h}$, $U_{h}^{\dagger}$, and $U_{v}^{\dagger}$ to search $U_{v}$ in a similar way by computing the matrix $M_{v}$ as follows,
\bea
\mysymbol{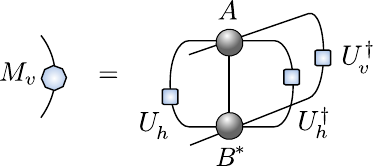}.
\eea 
We then obtain $U_{v} = V_{v}W_{v}^{\dagger}$ from $M_{v}= W_{v}S_{v}V_{v}^{\dagger}$.
\item[(v)] Compute the fidelity $F$ and check whether $F=1$, up to some acceptable small tolerance. If not, go back to step (i).
\end{itemize}

\section{Environment recycling in the TEBD algorithm for iMPS (1D) \label{secV}}

In this section we explain the environment recycling scheme in the TEBD algorithm for infinite 1D systems. For concreteness, we describe the scheme for the simple quantum Ising model with transverse magnetic field, where the time evolution operator is broken into gates representing the interaction and magnetic field. The infinite system described by this model can be represented by a one-site translationally invariant iMPS. The scheme can be generalized to arbitrary Hamiltonians and larger unit cells.

\subsection{Imaginary-time evolution for the 1D quantum Ising model \label{sec:ch5sec4sec1}}
Consider an infinite spin-$1/2$ chain described by the Hamiltonian of the quantum ferromagnetic Ising model defined as
\bea
H = -\sum_{i}\sigma^{z}_{i}\sigma^{z}_{i+1} - h\sum_{i}\sigma^{x}_{i},
\eea
where $\sigma^{z}, \sigma^{x}$ are Pauli matrices, $h$ is the amplitude of the transverse magnetic field. 

\begin{figure}[htpb]
\begin{center}
\includegraphics[width=\columnwidth]{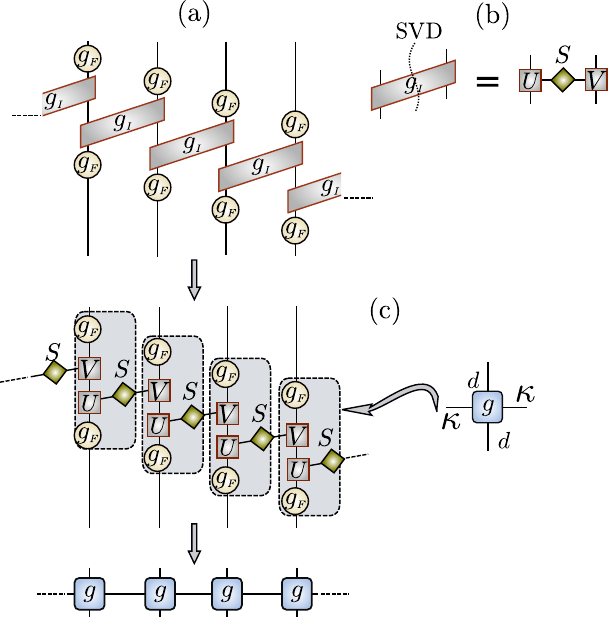}
\caption{ (Color online) Diagrammatic illustration for constructing an iMPO. (a) The imaginary-time evolution operator is decomposed into a product of nearest-neighbor interaction terms denoted by $g_{I}$ and magnetic field terms named $g_{F}$. (b) Perform the SVD of the interaction operator $g_{I} = USV$. (c) The iMPO is obtained by contracting all the tensors $g_{F}, U, V, S$ together.}
\label{fig15}
\end{center}\end{figure}

In order to find the ground state of the system employing the iTEBD algorithm, we firstly rewrite the Hamiltonian as
\bea
H = \sum_{i}H_{I}^{[i, i+1]} +\sum_{i}H_{F}^{[i]},
\eea
where $H_{I}^{[i, i+1]} =  -\sigma^{z}_{i}\sigma^{z}_{i+1}$ and $H_{F}^{[i]}= -h\sigma^{x}_{i}$ describe the nearest-neighbor interaction and magnetic field terms, respectively. The  commutation relations $[H_{I}^{[i, i+1]}, H_{I}^{[i', i'+1]}] = 0$ and $[H_{F}^{[i]}, H_{F}^{[i']}] = 0$ are satisfied whereas $[H_{I}^{[i, i+1]}, H_{F}^{[i']}] \neq 0$ possibly. The imaginary-time evolution operator at each time step $\delta$ can be decomposed by employing the second-order Suzuki-Trotter decomposition \cite{Suzuki1} such that
\bea
e^{-H\delta} &=& e^{-\sum_{i}H^{[i]}_{F}\delta/2}e^{-\sum_{i}H^{[i, i+1]}_{I}\delta}e^{-\sum_{i}H^{[i]}_{F}\delta/2} + O(\delta^3)\nn\\
&=&\prod_{i}g^{[i]}_{F}\prod_{\text{odd}~i}g^{[i, i+1]}_{I}\prod_{\text{even}~i}g^{[i, i+1]}_{I}\prod_{i}g^{[i]}_{F} + O(\delta^3),\nn\\
\eea
where we have defined
\bea
g^{[i]}_{F} =  e^{-H^{[i]}_{F}\delta/2}, ~~~~ g^{[i, i+1]}_{I} =  e^{-H^{[i, i+1]}_{I}\delta}.
\eea
As it is expected to find the ground state of the system that is represented by a one-site translationally invariant iMPS, we need to express the time evolution operator in terms of the matrix product operator (MPO) representation \cite{schollwock2,PirvuMPO,FrankMPO}. As illustrated in Fig.~\ref{fig15}, at each link between two lattice sites, we employ the SVD to decompose the interaction term as 
\bea
g^{[i, i+1]}_{I} = U^{[i]}S^{[i]}V^{[i+1]}.
\eea
with the same contribution between the interaction and magnetic field terms at each site, we define the MPO as follows,
\bea
g^{[i]} = g^{[i]}_{F}V^{[i]}S^{[i]}U^{[i]}g^{[i]}_{F}.
\label{iMPO1}
\eea

The \emph{infinite} MPO (iMPO) is then constructed by 
\bea
g_{\infty} = \bigotimes_{i}g^{[i]},
\eea
which is translationally invariant under a single rank-four tensor $g$ of dimensions $\kappa\times\kappa\times d\times d$. To obtain the ground state of the system represented by an iMPS, we apply successively the iMPO to an initial state $\ket{\Psi_0}$ until the convergence is achieved. Mathematically, this is described by
\bea
\ket{\Psi_{\GS}} &=& \lim_{T \rightarrow \infty} \frac{e^{-HT}\ket{\Psi_0}}{||e^{-HT}\ket{\Psi_0}||}\nn\\
&=&\lim_{m \rightarrow \infty} \frac{g^{m}\ket{\Psi_0}}{||g^{m}\ket{\Psi_0}||}\nn\\
&=&\lim_{m \rightarrow \infty} \frac{(\bigotimes_{i}g^{[i]})^{m}\ket{\Psi_0}}{||(\bigotimes_{i}g^{[i]})^{m}\ket{\Psi_0}||},
\label{eq}
\eea 
where $m = T/\delta$ and is the number of iterations in iTEBD algorithm.
\subsection{1D environment recycling\label{sec:ch5sec4sec2}}
We now describe the environment recycling scheme and how to incorporate it into the iTEBD algorithm to accelerate the convergence. We first need to determine quantities which characterize the environment in each update of the iTEBD algorithm. We assume that the system is represented by a canonical iMPS $\ket{\Psi_{A_{0}}}$ at time $t_{0}$ of the evolution. This iMPS is characterized by a tensor $A_{0}$ with bond dimension $\chi$. In the next update, at time $t_{0}+\delta$, the iMPS is updated to  $\ket{\Psi_{\Theta}}$ with bond dimension $\kappa\chi$ where $\Theta = gA_{0}$. The iMPS $\ket{\Psi_{\Theta}}$ is then transferred into its canonical to perform the bond dimension truncation to obtain a new iMPS $\ket{\Psi_{\tilde{A}}}$ with bond dimension $\chi$. This truncation is equivalent to finding matrices $P$ and $Q$ which dimensions are $\kappa\chi\times\chi$ and $\chi\times\chi\kappa$ respectively such that
\bea
\tilde{A} = Q\Theta P.
\label{EqCanoPQ1}
\eea
Note that the tensors $P$ and $Q$ are the objects that contain the information of the environment that we want to recycle. 

We assume that the environment characterized by tensors $P$ and $Q$ changes slightly for the next few time steps of the evolution, and thus these tensors can be recycled instead of being re-calculated from the scratch at every iteration. In order to propose the environment recycling scheme, we first need to construct a so-called \emph{renormalized gate} denoted as $G$ which is obtained by following the fundamental steps.
\begin{itemize}
\item[(i)] Transfer the iMPS $\ket{\Psi_{\tilde{A}}}$ into its canonical form, which is described by tensors $\{\tilde{\Gamma}_{1},\tilde{\lambda}_{1}\}$. We obtain new iMPS $\ket{\Psi_{\tilde{A}_{1}}}$ with $\tilde{A}_{1} = \sqrt{\tilde{\lambda}_{1}}\tilde{\Gamma}_{1}\sqrt{\tilde{\lambda}_{1}}$. There is also a relation between $\tilde{A}_{1}$ and $\tilde{A}$ such that $\tilde{A}_{1} = Q_{c}\tilde{A}P_{c}$, which is shown graphically as
\bea
\mysymbol{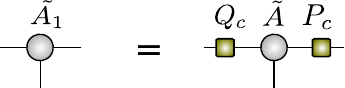},
\eea
 where $Q_{c}$ and $P_{c}$ are $\chi\times\chi$ matrices and can be obtained easily from the relation $Q_{c}\tilde{A}P_{c}= \sqrt{\tilde{\lambda}_{1}}\tilde{\Gamma}_{1}\sqrt{\tilde{\lambda}_{1}}$.
\item[(ii)] Assume that $\ket{\Psi_{A_{0}}}$ and $\ket{\Psi_{\tilde{A}_{1}}}$ represent a similar physical state, and hence, there is a unitary transformation such that $\tilde{A}_{1}\approx UA_{0}U^{\dagger}$ which is graphically shown as
\bea
\mysymbol{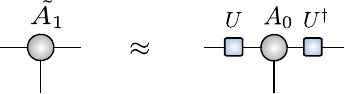},
\eea
where the unitary matrix $U$ can be found using the methods mentioned in Sec.~\ref{subsecIIIB}.
\item[(iii)] From the tensors $\{U,Q_{c},Q,g,P,P_{c}\}$, we construct a tensor $G$ such that $G = U^{\dagger}Q_{c}QgPP_{c}U$ which is visualized as
\bea
\mysymbol{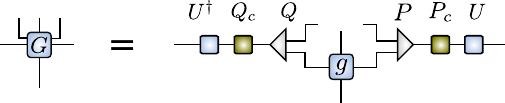}.
\eea
$G$ is called the renormalized gate.
\end{itemize}
If we apply the renormalized gate $G$ on the intial iMPS $\ket{\Psi_{A_{0}}}$, a new iMPS $\ket{\Psi_{A_{1}}}$ is obtained where $A_{1}$ has the same bond dimension with $A_{0}$ and the gauge degrees of freedom between them are fixed. At this point the environment recycling scheme can be applied by keeping applying the renormalized gate $G$ successively to tensor $A_{0}$ for several times to get a new state $\ket{\Psi_{A_{N}}}$ before computing a new environment for the next environment recycling process.
\subsection{1D benchmark results \label{sec:ch5sec4sec3}}
\begin{figure}[htpb]
\begin{center}
\includegraphics[width=\columnwidth]{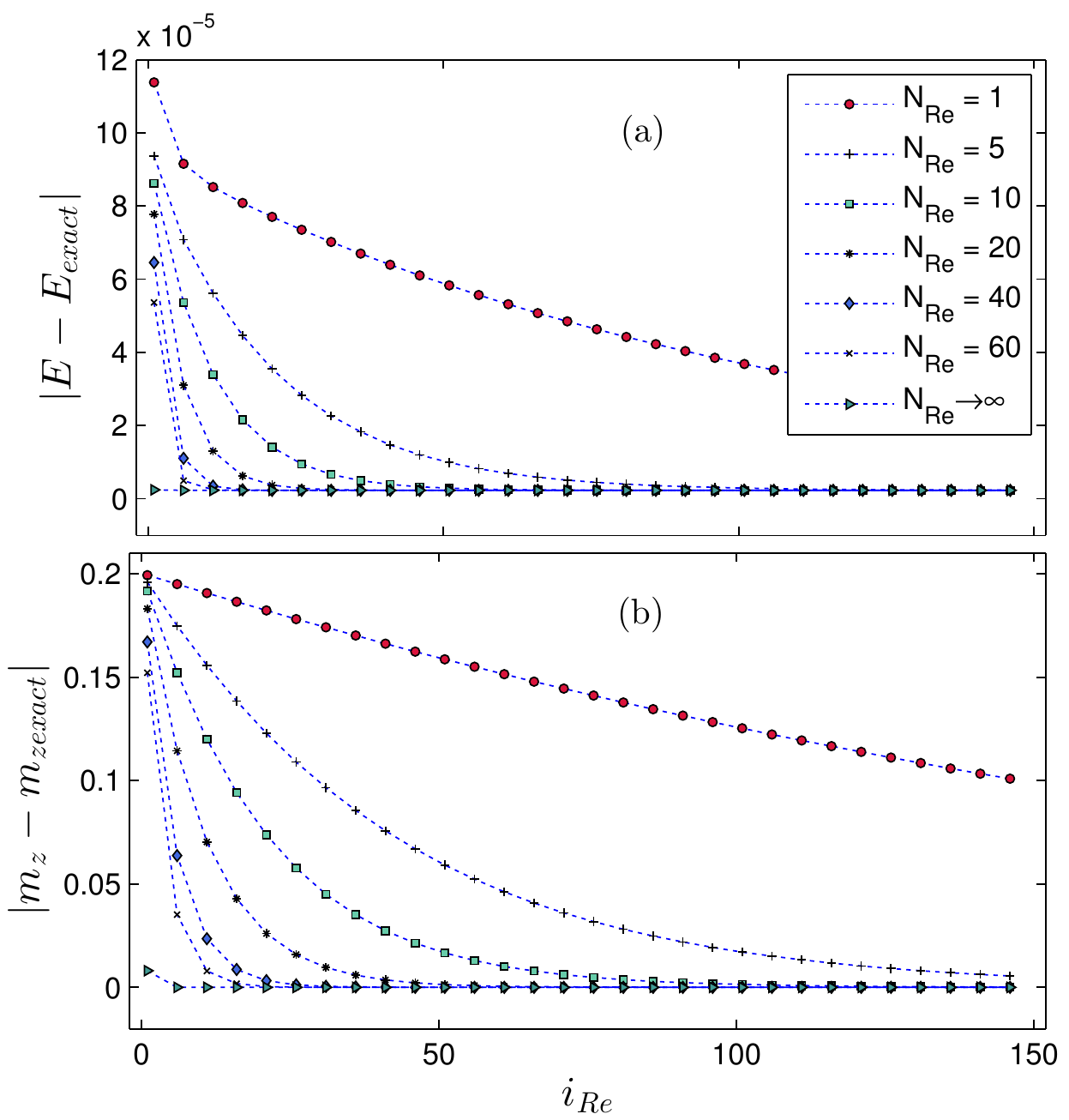}
\caption{ (Color online) Plots of relative errors in the local expectation values regarding the number of times $i_{Re}$ that the environment is calculated. The upper plot (a) shows the errors in the energy per link where $E_{exact} \approx -1.306856$. The bottom plot (b) shows the errors in the local magnetization per site where $m_{z_{exact}}=0$. The results for different $N_{Re}$'s are obtained by simulating with the same initial iMPS of bond dimension $\chi = 50$, $i_{Re}$ is counted from one, the transverse magnetic field $h=1.05$, time step $\delta = 0.05$. The $N_{Re}\rightarrow\infty$ lines correspond to the cases where the dominant eigenvalue of the renormalized gate is found.}
\label{fig19}
\end{center}\end{figure}

We have incorporated the environment recycling scheme into the iTEBD algorithm for iMPS to study the ground state of the 1D Ising model with the transverse magnetic field $h = 1.05$. In Fig.~\ref{fig19}, we show the relative errors during the convergence of the ground state energy per link and local magnetization per site versus the number of times that the environment is calculated, denoted as $i_{Re}$, for different numbers of recyclings $N_{Re}$'s. The total number of time steps is $N_{ts} = N_{Re}i_{Re}$. The larger the $N_{ts}$ the closer to the ground state. Therefore, in the plots, we see that with the same $i_{Re}$, the larger the $N_{Re}$ smaller the relative error. Besides, note that unless $N_{Re}$ is extremely large, it may take the same number of time steps $N_{ts}$ to obtain the same ground state of the system for different $N_{Re}$'s. However, when $N_{Re}> 1$, the computational effort is already reduced because the number of environment calculations is reduced. Therefore, the larger the $N_{Re}$, the faster the ground-state convergence.

A question naturally arises: can we keep recycling a particular environment as many times as we want? In principle, the answer is yes and the maximum of $N_{Re}$ can be reached when the dominant eigenvector of the renormalized gate is found. Roughly, this can be understood as recycling the environment with an infinite number of times, i.e., $N_{Re}\rightarrow\infty$.  However, if $N_{Re}$ is chosen to be very large, the scheme can be unstable as the state of the system might fall into the local minima. Therefore, it depends on the problem, we can perform numerical calculation to choose an appropriate value for $N_{Re}$.

\section{Environment recycling in the TEBD algorithm for iPEPS (2D) \label{secVI}}

In the previous section we have shown that the environment recycling scheme works quite well for 1D systems using iMPS. Although it is not a crucial improvement of the TEBD with iMPS with OBC (recycling the environment in that case does not reduce significantly the computational cost), it is a good start to build the environment recycling scheme for 2D systems. In this section, we will present in detail a scheme for 2D infinite systems using iPEPS. For concreteness, we propose the scheme to study the ground state of an infinite square lattice system described by the 2D quantum Ising model with a transverse magnetic field.

\subsection{Imaginary-time evolution for the 2D quantum Ising model \label{sec:ch5sec5sec1}} 
The ferromagnetic Ising Hamiltonian for a quantum lattice system in 2D has the following form
\bea
H = -\sum_{<\vec{r},\vec{r}'>}\sigma^{z}_{\vec{r}}\sigma^{z}_{\vec{r}'} - h\sum_{\vec{r}}\sigma^{x}_{\vec{r}},
\eea
where $<\vec{r},\vec{r}'>$ represents the nearest-neighbor sites. For a square lattice, each lattice site has four nearest neighbors. We then group the nearest-neighbor interactions into two terms, for interactions along the horizontal and vertical directions as follows,
\bea
H = -\sum_{i,j}\sigma^{z}_{(i,j)}\sigma^{z}_{(i+1,j)} - \sum_{i,j}\sigma^{z}_{(i,j)}\sigma^{z}_{(i,j+1)} - h\sum_{i,j}\sigma^{x}_{(i,j)},\nn\\
\eea
where we have used indices $i$ and $j$ to represent horizontal and vertical directions in the square lattice respectively, i.e., $\vec{r} = (i,j)$. The Hamiltonian can be also written as 
\bea
H = \sum_{i,j}H_{Ih}^{[i,i+1]j}+ \sum_{i,j}H_{Iv}^{i[j,j+1]} +\sum_{i,j}H_{F}^{[i,j]},
\eea
where $H_{Ih}^{[i,i+1]j} =  -\sigma^{z}_{(i,j)}\sigma^{z}_{(i+1,j)}$ and $H_{Iv}^{i[j,j+1]} =  -\sigma^{z}_{(i,j)}\sigma^{z}_{(i,j+1)}$ are the horizontal and vertical nearest-neighbor interactions, respectively, and $H_{F}^{[i,j]}= -h\sigma^{x}_{(i,j)}$ is the one-site magnetic field. Notice that the commutation relations $[H_{Ih}^{[i,i+1]j}, H_{Ih}^{[i',i'+1]j'}] = 0, [H_{Iv}^{i[j,j+1]}, H_{Iv}^{i'[j',j'+1]}] = 0, [H_{Ih}^{[i,i+1]j}, H_{Iv}^{i'[j',j'+1]}] = 0$ and $[H_{F}^{[i,j]}, H_{F}^{[i',j']}] = 0$ are satisfied whereas in general $[H_{Ih}^{[i,i+1]j}, H_{F}^{[i',j']}] \neq 0$, $[H_{Iv}^{i[j,j+1]}, H_{F}^{[i',j']}] \neq 0$. 

\begin{figure}[htpb]
\begin{center}
\includegraphics[width=\columnwidth]{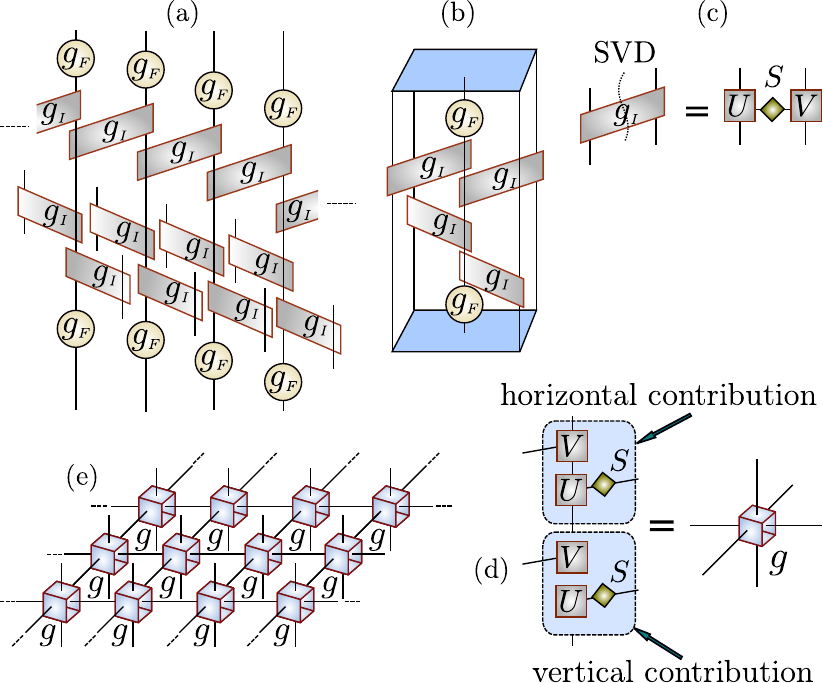}
\caption{ (Color online) Diagrammatic illustration for building an iPEPO. (a) The imaginary-time evolution operator is decomposed into two terms: the interaction terms denoted by $g_{I}$ in horizontal and vertical directions and the on-site magnetic field terms $g_{F}$. (b) A PEPO will be created from the horizontal and vertical contributions and the on-site magnetic field terms. (c) Perform the SVD of the interaction operators $g_{I} = USV$. (d) PEPO obtained from contracting tensors $g_{F}, U, V, S$. Each PEPO is a tensor consisting of two physical indices of dimension $d$ and four virtual indices ofdimension $\kappa$. (e) An iPEPO.}
\label{fig20}
\end{center}\end{figure}

In order to obtain the ground state of the system, we again employ the imaginary-time evolution. To start, we need to express the imaginary-time evolution operator in terms of the projected entangled-pair operator (PEPO). This can be done by decomposing the imaginary-time evolution operator as follows: 
\begin{widetext}
\bea
e^{-H\delta} &=& e^{-\sum_{i,j}H_{F}^{[i,j]}\delta/2}e^{-\sum_{i,j}H_{Ih}^{[i,i+1]j}\delta}e^{-\sum_{i,j}H_{Iv}^{i[j,j+1]}\delta}e^{-\sum_{i,j}H_{F}^{[i,j]}\delta/2} + O(\delta^3)\nonumber\\
&=&\prod_{i,j}g^{[i,j]}_{F}\prod_{j,~\text{odd}~i}g^{[i,i+1]j}_{Ih}\prod_{j,~\text{even}~i}g^{[i,i+1]j}_{Ih}\prod_{i,~\text{odd}~j}g^{i[j,j+1]}_{Iv}\prod_{i,~\text{even}~j}g^{i[j,j+1]}_{Iv}\prod_{i,j}g^{[i,j]}_{F} + O(\delta^3),\nn\\
\eea
\end{widetext}
where we have defined
\bea
g^{[i,j]}_{F} &=&  e^{-H^{[i,j]}_{F}\delta/2}, g^{[i,i+1]j}_{Ih} =  e^{-H^{[i,i+1]j}_{Ih}\delta}, \nn\\
g^{i[j,j+1]}_{Iv} &=&  e^{-H^{i[j,j+1]}_{Iv}\delta}.\nn
\eea
We construct the PEPO in a similar way with what we did above for the MPO. The scheme is explained diagrammatically in Fig.~\ref{fig20}. More concretely, we apply SVD to decompose the interaction components in both horizontal and vertical directions of the lattice as follows:
\bea
g^{[i,i+1]j}_{Ih} = U^{[i,j]}_{h}S^{[i,j]}_{h}V^{[i+1,j]}_{h}
\eea
and 
\bea
g^{i[j,j+1]}_{Iv} = U^{[i,j]}_{v}S^{[i,j]}_{v}V^{[i,j+1]}_{v},
\eea
respectively; see Fig.~\ref{fig20}(c). With the equal contributions of the interaction terms (from both directions) and magnetic field term, the PEPO at each lattice site is constructed by
\bea
g^{[i,j]} = g^{[i,j]}_{F}V^{[i,j]}_{h}U^{[i,j]}_{h}S^{[i,j]}_{h}V^{[i,j]}_{v}U^{[i,j]}_{v}S^{[i,j]}_{v}g^{[i,j]}_{F},
\eea
as illustrated in Fig.~\ref{fig20} (d). In Fig.~\ref{fig20} (e), the iPEPO is then defined as a tensor product of identical local terms, which is given by 
\bea
g_{\infty} = \bigotimes_{i,j}g^{[i,j]}.
\eea

The ground state of system will be obtained by acting  on the iPEPS with the iPEPO gate as many times as needed until it is converged. We keep in mind that, as a result of acting the iPEPO gate on the iPEPS at each update, the bond dimension of this iPEPS will also increase by $\kappa$. Thus, one needs to perform a truncation so that the bond dimension will remain the same after each update and the obtained iPEPS is also the best approximated ground state. At this point, we can take advantage of transferring the iPEPS into its canonical form to perform the truncation. This scheme is also known as the SU \cite{JiangSimplify1,LiSimplify1,Kalis1,RomanNewCanonical1}, applied here for a one-site translationally invariant system. The advantage of the SU is obvious as the computational cost is reduced essentially as compared to the FU. However, in most of the cases, results obtained from this scheme are less accurate, especially when applied to a critical system where the correlations are long-range. In these cases, we need the FU, which involves calculating the environment and will be explained next.
\subsection{Full update for a one-site translationally invariant iPEPS \label{sec:ch5sec4sec2}}
The FU is typically implemented using two-site translationally invariant iPEPS \cite{Jordan1, Roman2, Philippe4}. In what follows, we will describe the FU for a one-site translationally invariant iPEPS. 

At each time step of the evolution, after applying the iPEPO gate to the iPEPS, we get a new state $\ket{\Psi_{g}} = g_{\infty}\ket{\Psi_{A}}$ which is represented by tensor $\Theta$ with bond dimension $\kappa D$. We need to find four optimal tensors $\{P_{h}, Q_{h},P_{v}, Q_{v}\}$ to insert into four bonds of the tensor $\Theta$ simultaneously such that the updated iPEPS $\ket{\Psi_{\tilde{A}}}$ represented by $\tilde{A}= Q_{h}Q_{v}\Theta P_{h}P_{v}$ with bond dimension $D$ is well-approximated to $\ket{\Psi_{g}}$ and this can be achieved by employing the FU.

The FU involves the two following fundamental processes.
\begin{enumerate}
 \item \emph{Environment calculation:}

There are a lot of feasible ways to compute the environment. For instance, one can utilize the coarse-graining tensor renormalization group methods, like TRG/SRG and HOTRG/HOSRG \cite{Levin1,Xie1,Zhao1,Xie2}. Alternatively, the environment can be also computed by means of the iTEBD algorithm applied to a boundary iMPS\cite{Jordan1}. Here we apply the corner transfer matrix method (CTM) \cite{Baxter11,Nishino1,Nishino1_1,Roman2,Philippe4,RomanNewCTM1} to compute the environment. This method is known to be more efficient than the MPS schemes when studying the system in the vicinity of the phase transition \cite{Roman2}. In our case, an infinite square lattice $\mathcal{L}(a)$,  where $a_{(l_{1}l_{2}),(r_{1}r_{2}),(d_{1}d_{2}),(u_{1}u_{2})} = \sum_{s}\Theta^{s}_{l_{1}r_{1}d_{1}u_{1}}(\Theta^{s}_{l_{2}r_{2}d_{2}u_{2}})^{*}$,  is constructed by contracting the physical indices of  $\ket{\Psi_{g}}$ and $\bra{\Psi_{g}}$. We decompose $\mathcal{L}(a)$ into two contiguous regions $\mathcal{A}$ and $\mathcal{B}$; see Fig.~\ref{fig21}(a). $\mathcal{B}$ is chosen to contain four sites where the tensors are connected to each other by the links denoted as \emph{r} (right), \emph{d} (down), \emph{l} (left), and \emph{u} (up). $\mathcal{A}$ plays a role as an environment around $\mathcal{B}$ and is required to be computed. Applying the CTM method, this environment is computed  approximately and characterized by a set of four corner tensors $\{C_{1}, C_{2}, C_{3}, C_{4}\}$ and eight edge tensors $\{Ta_{1}, Ta_{2}, Ta_{3}, Ta_{4}, Tb_{1}, Tb_{2}, Tb_{3}, Tb_{4}\}$, see Fig.~\ref{fig21}(a). Note that we can further contract the network to obtain the environment represented by only six tensors $\{E_{1}, E_{2}, E_{3}, E_{4}, E_{5}, E_{6}\}$ around a specific link, see Fig.~\ref{fig21}(b). 
\begin{figure}[htpb]
\begin{center}
\includegraphics[width=\columnwidth]{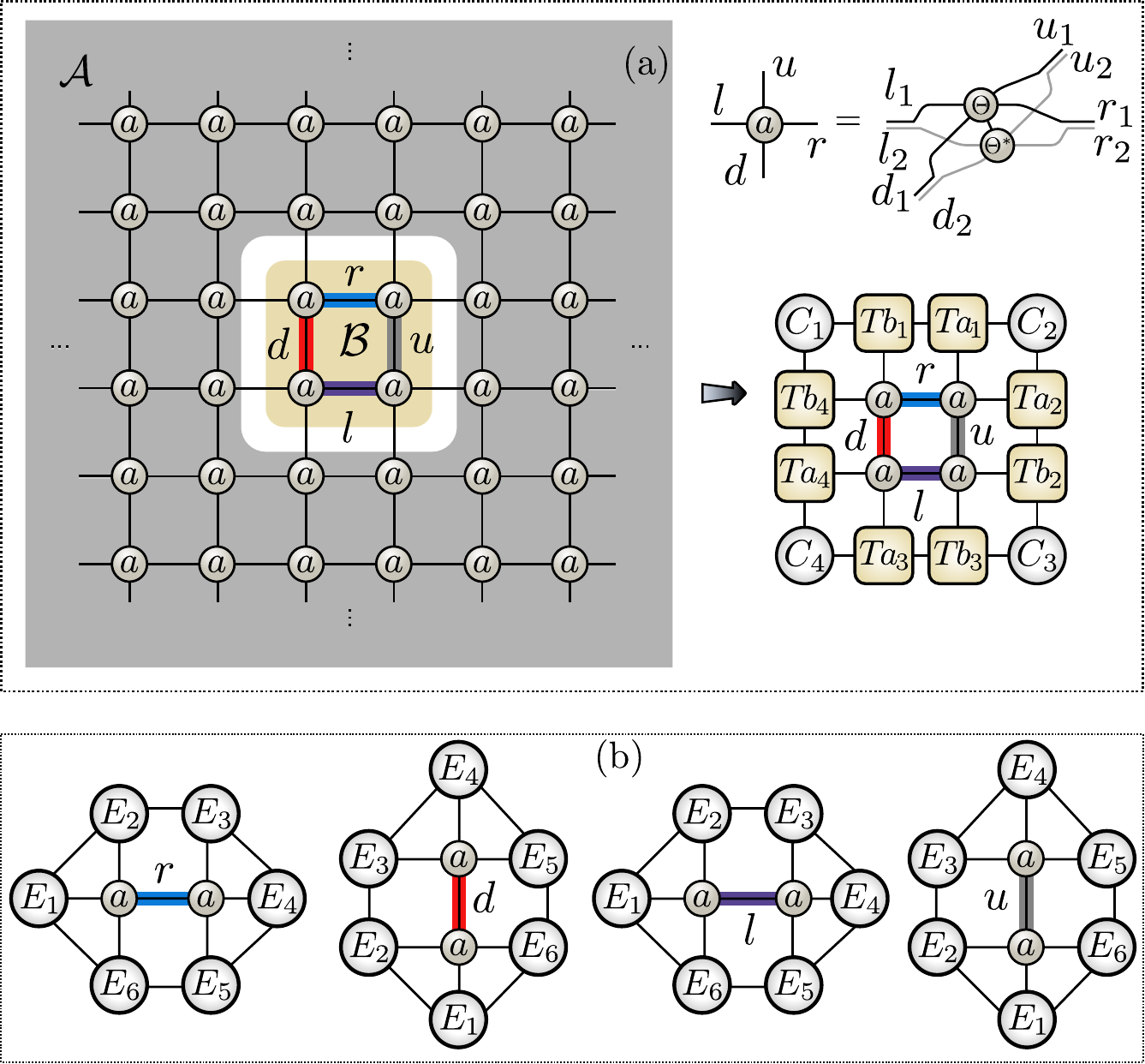}
\caption{ (Color online) Compute the environment around a specific link using the CTM method. (a) An infinite square lattice $\mathcal{L}(a)$ is decomposed into two contiguous parts $\mathcal{A}$ and $\mathcal{B}$. The environment $\mathcal{A}$ around $\mathcal{B}$  is characterized by corner tensors $\{C_{1}, C_{2}, C_{3}, C_{4}\}$ and edge tensors $\{Ta_{1}, Ta_{2}, Ta_{3}, Ta_{4}, Tb_{1}, Tb_{2}, Tb_{3}, Tb_{4}\}$. (b) The environment around a specific link is represented by six tensors $\{E_{1}, E_{2}, E_{3}, E_{4}, E_{5}, E_{6}\}$.}
\label{fig21}
\end{center}\end{figure}

In order to obtain optimal tensors  $\{P_{h}, Q_{h},P_{v}, Q_{v}\}$, we need to variationally optimize the pairs of tensors  $(P_{r}, Q_{r})$, $(P_{d}, Q_{d})$, $(P_{l}, Q_{l})$, and $(P_{u}, Q_{u})$ so that they optimally truncate the bond dimensions of the tensor $\Theta$ corresponding to the links \emph{ r, d, l,} and \emph{u}, respectively. This can be done via the standard variational method explained below.

\item \emph{ Standard variational method:} 
Let us define a cost function as follows:
\bea
f(\zeta)&=&||\ket{\Psi_{g}}-\ket{\Psi_{\zeta}}||^{2}\nn\\
&=&\braket{\Psi_{g}}{\Psi_{g}} + \braket{\Psi_{\zeta}}{\Psi_{\zeta}}-\braket{\Psi_{g}}{\Psi_{\zeta}}- \braket{\Psi_{\zeta}}{\Psi_{g}},\nn\\
\label{var}
\eea   
where $\zeta = (P_{r}, Q_{r}, P_{d}, Q_{d},P_{l}, Q_{l},P_{u}, Q_{u})$. Note that $f(\zeta)$ is a quadratic function for every variables in $\zeta$, and hence we can use a standard variational method to find $\zeta$ that minimizes $f(\zeta)$. 
More concretely, let us fix all tensors in $\zeta$ except for the first two tensors $(P_{r}, Q_{r})$. We then find $P_{r}$ and $Q_{r}$ that minimize the cost function by following the steps.
\begin{itemize}
 \item[(i)] Fix $Q_{r}$ with some arbitrary initial tensor or the tensor obtained from previous iteration and find $P_{r}$. Then we can write  Eq.~(\ref{var}) as a quadratic scalar expression as follows,
\bea
f(P_{r},P_{r}^{\dagger})&=&P_{r}^{\dagger}RP_{r}-P_{r}^{\dagger}S-S^{\dagger}P_{r}+T,
\label{var1}
\eea
where $P_{r}$ is understood as a reshaped vector with $D^{2}\kappa$ components and matrices $R, S, T$ can be obtained from contracting appropriate TNs including the contraction of the whole environment around the link \emph{r}.
\item[(ii)] From Eq.~(\ref{var1}), the minimum of $f(P_{r},P_{r}^{\dagger})$ with respect to $P_{r}^{\dagger}$ is obtained only if $P_{r} =R^{-1} S$. 
\item[(iii)] Fix the tensor $P_{r}$ and find $Q_{r}$ with the same procedure as steps (i) and (ii) above.
\end{itemize}
\end{enumerate}
Update the new tensors $P_{r}$ and $Q_{r}$ in $\zeta$ and then turn into the links \emph{d, l} and \emph{u} to update $(P_{d}, Q_{d})$, $(P_{l}, Q_{l})$, and $(P_{u}, Q_{u})$, respectively.

The above process are iterated until all the tensors in $\zeta$ converge to fixed tensors. The convergence can be recognized by tracing of the cost function computed in the form of Eq.~(\ref{var}).  Note also that  instead of applying the optimization scheme described above one could also apply other methods such as the conjugate gradient algorithm to optimize these tensors.

Once the tensors $(P_{r}, Q_{r}, P_{d}, Q_{d},P_{l}, Q_{l},P_{u}, Q_{u})$ are converged, we can set $P_{h} = (P_{r}+P_{l})/2$, $Q_{h} = (Q_{r}+Q_{l})/2$, $P_{v} = (P_{d}+P_{u})/2$ and $Q_{v} = (Q_{d}+Q_{u})/2$. This is due to the links \emph{r} and \emph{l} are equivalent and so are the links \emph{d} and \emph{u}. The new iPEPS is represented by $\tilde{A}= Q_{h}Q_{v}\Theta P_{h}P_{v}$. 

If the state is evolved by a large number of time steps, where at each time step one  apply the iPEPO to the iPEPS combined with the bond dimension truncation using the above FU scheme, a converged iPEPS is obtained which represents approximately the ground state of the system.

Note that the computational cost of this FU is quite expensive, as the cost roughly scales as $\mathcal{O}(\chi^{3}D^{6}\kappa^{6})$. This dominant cost is due to the environment calculation. Compared to the FU for a two-site translationally invariant iPEPS, where the dominant cost is $\mathcal{O}(\chi^{3}D^{6})$ as stated in Ref.~\onlinecite{Roman2}, this one-site translationally invariant scheme is more expensive by a factor $\kappa^{6}$, and hence, is much less efficient. Fortunately, we can get rid of the factor $\kappa^{6}$ by truncating the tensor $\Theta$ before sandwiching it to tensor $a$ to calculate the environment. The truncation can be performed by using tensors $\{P_{h}, Q_{h},P_{v}, Q_{v}\}$ which can be obtained from the previous update or the SU. In our simulation we use the latter and it works very well. With this optimization, we reduce the computational cost down to $\mathcal{O}(\chi^{3}D^{6})$ which is the same as for the case of using two-site translationally invariant iPEPS.

\subsection{2D environment recycling scheme\label{sec:ch5sec5sec3}}

The above FU provides a method to study the ground state of a one-site translationally invariant iPEPS, with the same computational cost as the usual two-site algorithm. However, this is still computationally very expensive and more importantly it requires a long time to converge to the ground state. The main aim of this paper is to show how we can further improve this drawback. This can be done  by incorporating the environment recycling scheme into the iTEBD algorithm for iPEPS to speed up the convergence.

The 2D recycling environment scheme is constructed similarly with the case of 1D. More precisely, suppose that at time $t_{0}$ of the imaginary-time evolution, the iPEPS $\ket{\Psi_{A_{0}}}$ is in the canonical form and characterized by tensor $A_{0}$ with bond dimension $D$. At time $t_{0}+\delta$, we apply the iPEPO to this iPEPS by contracting tensor $g$ with $A_{0}$. This results in tensor $\Theta$ of bond dimension $\kappa D$. To keep the bond dimension of the iPEPS fixed at $D$, we employ the FU described above to get a new tensor $\tilde{A}= Q_{h}Q_{v}\Theta P_{h}P_{v}$, which is visualized as:
\bea
\mysymbol{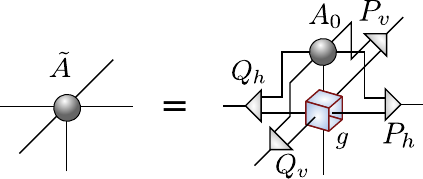}.
\eea
Note that the tensors $\{Q_{h},Q_{v}, P_{h},P_{v}\}$ indirectly contain the environment. As the environment is assumed to change slightly for the next few time steps of the evolution, we can recycle them. Similar to the case 1D, the recycling environment scheme includes two phases: define the \emph{renormalized gate} $G$ and successively apply $G$ a few time steps on $A_{0}$. In order to determine $G$, we proceed with the following fundamental steps:
\begin{itemize}
\item[(i)] Transfer the iPEPS $\ket{\Psi_{\tilde{A}}}$ into its canonical form, which is charaterized by tensors $\{\tilde{\Gamma},\tilde{\lambda}_{h},\tilde{\lambda}_{v}\}$. We then assign $\tilde{A}_{1} =\sqrt{\tilde{\lambda}_{h}}\sqrt{\tilde{\lambda}_{v}}\tilde{\Gamma}\sqrt{\tilde{\lambda}_{h}}\sqrt{\tilde{\lambda}_{v}}$ for the iPEPS $\ket{\Psi_{\tilde{A}_{1}}}$. Next, we can easily find the matrices $\{Q_{hc},Q_{vc},P_{hc},P_{vc}\}$ such that $\tilde{A}_{1} = Q_{hc}Q_{vc}\tilde{A}P_{hc}P_{vc}$ which can be seen graphically as,
\bea
\mysymbol{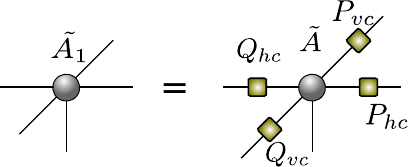}.
\eea
\item[(ii)] Assume that $\ket{\Psi_{\tilde{A}}}$ and $\ket{\Psi_{\tilde{A}_{1}}}$ represent a similar physical state, then we can find the unitary matrices to fix the gauge between two tensors $A_{0}$ and $\tilde{A}_{1}$ such that $\tilde{A}_{1}\approx U_{h}U_{v}A_{0}U_{h}^{\dagger}U_{v}^{\dagger}$ which is visualized as,
\bea
\mysymbol{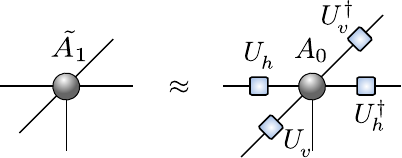}.
\eea
\item[(iii)] Construct the renormalized gate defined as $G = U_{h}^{\dagger}U_{v}^{\dagger}Q_{hc}Q_{h}Q_{vc}Q_{v}gP_{h}P_{hc}Q_{v}P_{v}P_{vc}U_{h}U_{v}$ which is graphically shown as follows,
\bea
\mysymbol{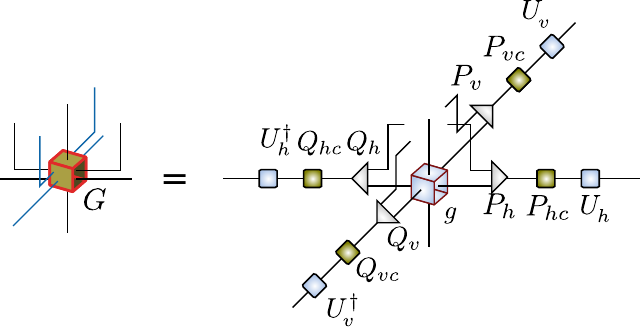}.
\eea
\end{itemize}
The recycling process is implemented by applying successively the renormalized gate $G$ to state $\ket{\Psi_{A_{0}}}$ for several times to get a new state $\ket{\Psi_{A_{N}}}$. Next, we re-assign state $\ket{\Psi_{A_{0}}}=\ket{\Psi_{A_{N}}}$ before applying another environment recycling process.
\subsection{2D benchmark results \label{sec:ch5sec5sec4}}
\begin{figure}[htpb]
\begin{center}
\includegraphics[width=\columnwidth]{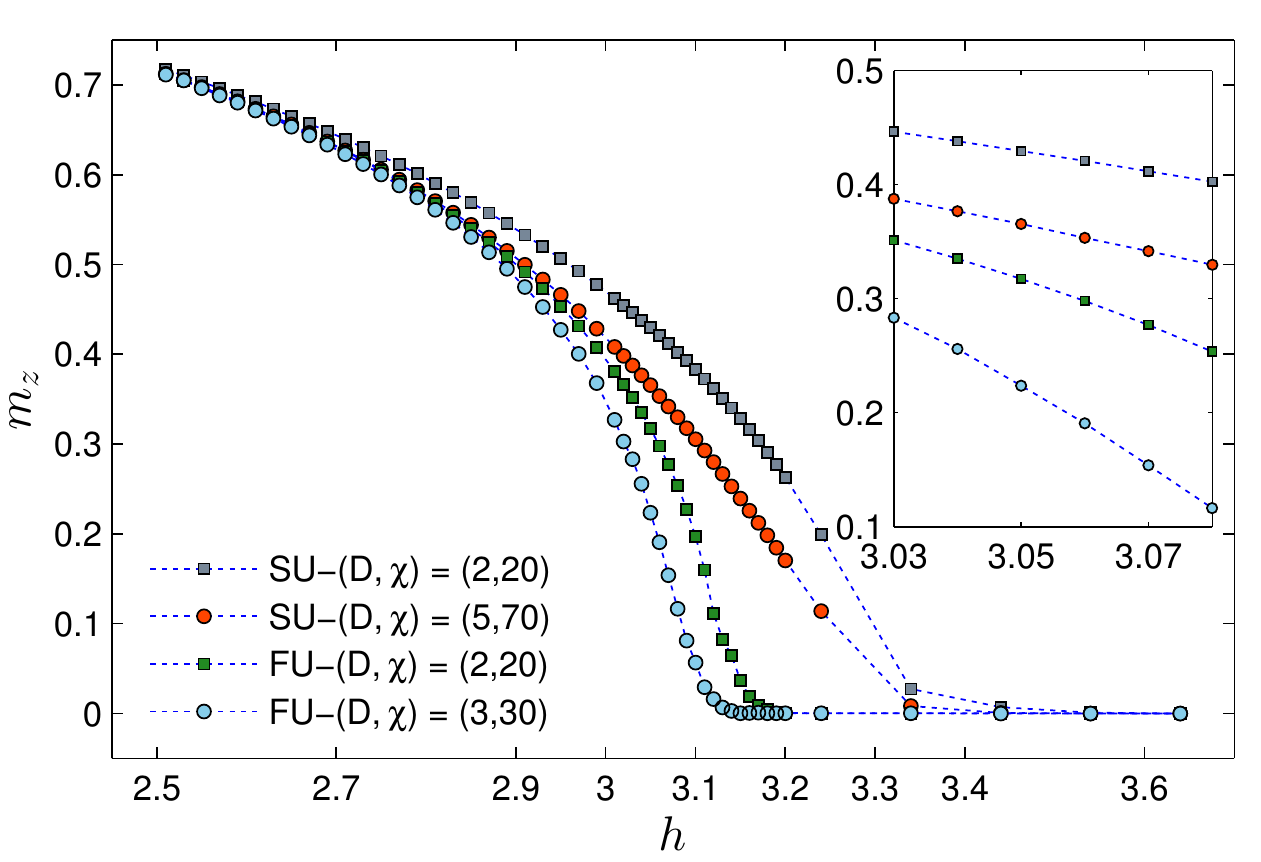}
\caption{ (Color online) Expectation value for the local magnetization per site $m_{z}$  as a function of the transverse magnetic field $h$. The result is obtained by using a one-site translationally invariant iPEPS simulating an imaginary time evolution with the iTEBD algorithm, both with SU and FU. The results close to the criticality are shown in the inset.}
\label{fig26}
\end{center}\end{figure}

In Fig.~\ref{fig26}, we plot the local magnetization $m_z(h)$ per site, which is defined as
\bea
m_{z}(h) = \frac{\braket{\Psi_{h}}{\sigma^{z}|\Psi_{h}}}{\braket{\Psi_{h}}{\Psi_{h}}},
\eea
 where $\ket{\Psi_{h}}$ is the ground state pertaining to the magnetic field $h$. The plot also shows the comparisons between different update schemes: SU and FU. We can see that the result obtained from the FU are much better than the results of SU, especially in the vicinity of the criticality.

\begin{figure}[htpb]
\begin{center}
\includegraphics[width=\columnwidth]{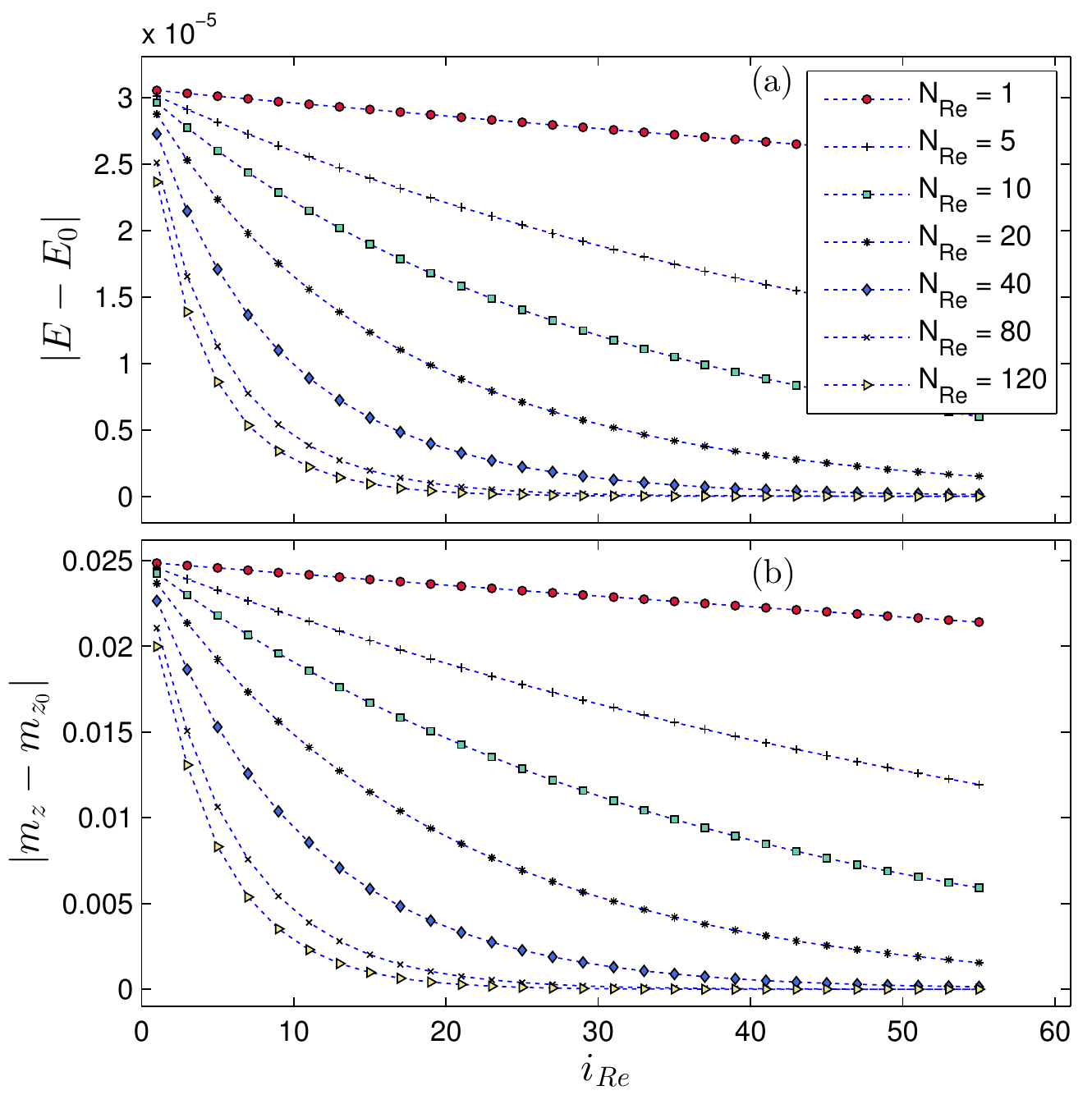}
\caption{ (Color online) Plots of relative errors in the local expectation values regarding the number of times $i_{Re}$ that the environment is calculated. The upper plot (a) shows the errors in the energy energy per link where $E_{0} \approx -1.619581$. The bottom plot (b) shows the errors in the local magnetization per site where $m_{z_{0}} \approx  0.240717$. The results for different $N_{Re}$'s are obtained by simulating with the same initial iPEPS of bond dimension $D=3$, $\chi=30$, $i_{Re}$ is counted from one, the transverse magnetic field $h=3.05$ and time step $\delta = 0.005$.}
\label{fig27}
\end{center}
\end{figure}

\begin{figure}[htpb]
\begin{center}
\includegraphics[width=\columnwidth]{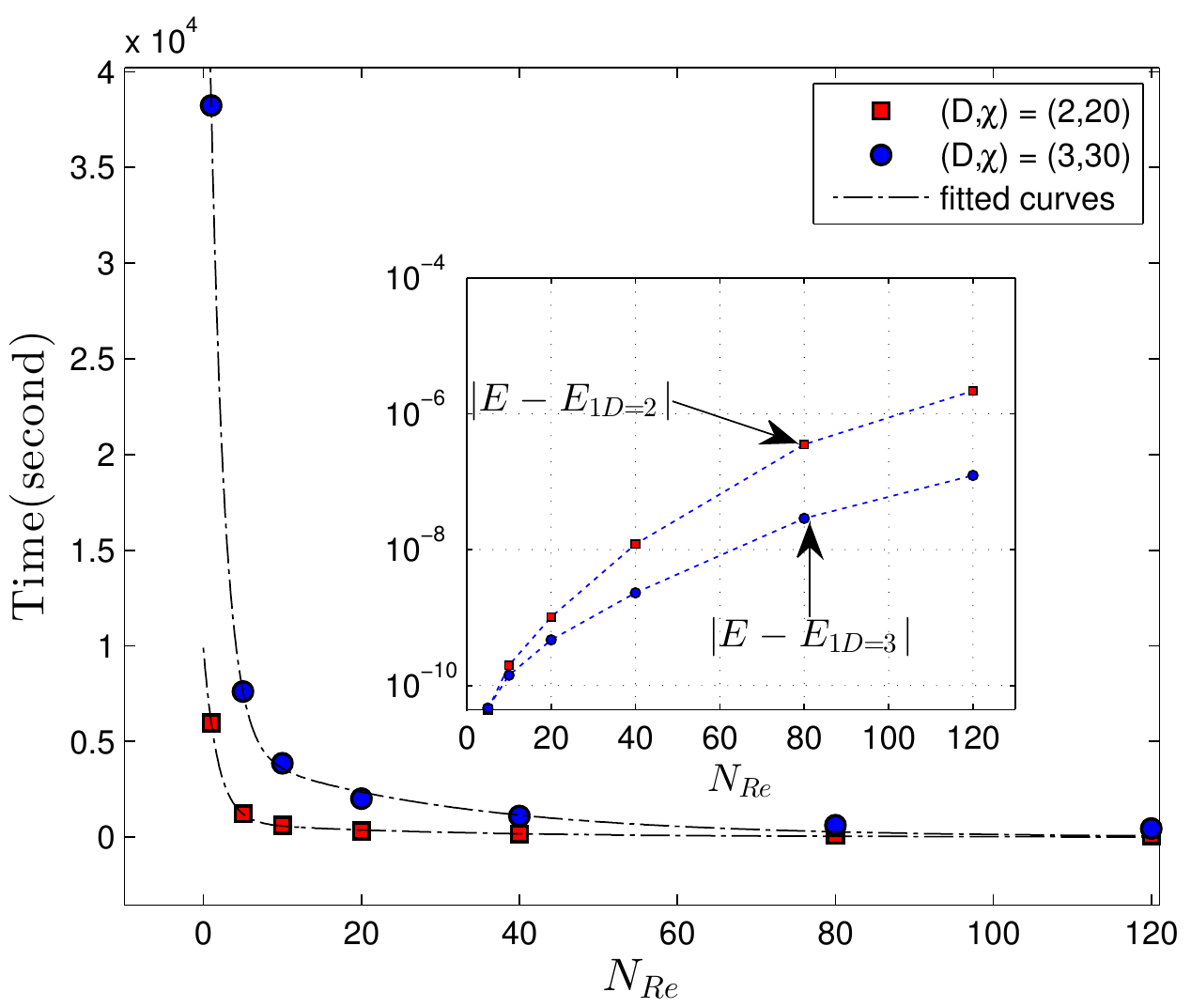}
\caption{ (Color online) Plots of the elapsed time measured in terms of the number of environment recyclings $N_{Re}$. The system is evolved with the same amount of time $T=12.05$ at time step $\delta = 0.01$ and $h = 3.01$. The plots are shown for different iPEPS bond dimensions where the exponential fitting curves are added to observe the abrupt decay in computational time when $N_{Re}$ increases. The inset shows the relative errors in the energy per link obtained after $T$ compared between $N_{Re}=1$ and the others in each case of bond dimension where $E_{1D=2}\approx-1.601426$ and $E_{1D=3}\approx-1.601809$.}
\label{fig28}
\end{center}
\end{figure}

Results of environment recycling scheme for the 2D quantum Ising model are shown in Fig.~\ref{fig27} using iPEPS with bond dimensions $D = 3$ and $\chi=30$ is used for computing the environment with CTM. We observe how fast the algorithm drives the system to the ground state by calculating the energy per link $E(h)$ and magnetization per site $m_{z}(h)$, with different numbers of environment recyclings near the critical point $h = 3.05$. In both plots we realize that the larger the number of environment recyclings the faster the convergence into ground state we obtain. We show extra plots in Fig.~\ref{fig28} to compare the time consumed when the system is evolved with the same amount of computational time $T = N_{ts}\delta$ using different numbers of environment recyclings. The plots are illustrated for different bond dimensions of the iPEPS, i.e., $(D,\chi) = (2,20)$ and $(D,\chi) = (3,30)$. The plots show that the larger the $N_{Re}$, the smaller the time elapsed to evolve the system up to time $T$. We also fit the data and observe that the time measured almost decays exponentially with the number of environment recyclings.

Note also that as in the case of 1D environment recycling scheme, it is not always allowed to use a freely large number of recyclings although the larger it is the faster convergence to ground state. This is also applied for the case 2D.  The reason is that instability appears if the same environment used to update the state that might be significantly different from the initial state, and hence it is contrary to the assumption that the states between successive time evolution update steps are close enough to each other. Therefore, we need to wisely choose an appropriate number of recyclings to stablize the algorithm when dealing with a specific problem.

\section{Conclusions\label{secVII}}

In this paper, we have first introduced a canonical form for both MPS (with PBC) in 1D and PEPS in 2D, as the fixed point of an iterative procedure, and used this canonical form as the basis to completely fix the gauge freedom in the MPS and PEPS. Then we have introduce a strategy to recycle the environment during a simulation of imaginary time evolution using the TEBD algorithm with iMPS and iPEPS.

By recycling the environment, we have obtained a significant reduction of computational costs while using the FU approach. The speed-up is accomplished because we only need to compute the environment once and can apply it for several time steps of the imaginary time evolution. To validate the schemes, we have studied the ground state of the quantum Ising model with transverse magnetic field in 1D and 2D and shown that the convergence into ground state is greatly accelerated the computational time required to evolving the system for a given amount of imaginary time decays abruptly. However, the incorporation of environment recycling scheme into the TEBD algorithm with iPEPS sometimes causes numerical instability. For instance, if the initial state on which the time evolution is applied is very different than the ground state of the system, we find it is best to first use a FU scheme without recycling of the environment for a few time steps, and only use recycling once the many-body state has stopped changing dramatically at each time step. Similarly, the number $N_{Re}$ of time steps during which the environment is recycled needs to be adjusted to ensure that the environment does not become too out-dated. 

In summary, we have proposed an environment recycling strategy that significantly accelerates ground state computations using iPEPS. We envisage that this strategy will help boost the application of PEPS to challenging 2D problems, such as the resolution of \emph{t-J}  and Hubbard models \cite{Philippe3,Philippe5}.

{\it Acknowledgment-} This research was supported in part by the ARC Centre of Excellence in Engineered Quantum Systems (EQuS), Project No. CE110001013 and the Discovery Projects funding scheme, Project No. DP1092513. H. N. Phien would like to thank the Centre for Health Technologies at The University of Technology Sydney for hospitality when completing this research. G.Vidal thanks the EQuS for hospitality at the University of Queensland, the NSERC (discovery grant), the John Templeton Foundation (Emergence grant) and the Simons Foundation (Many Electron Collaboration). Research at Perimeter Institute is supported by the Government of Canada through Industry Canada and by the Province of Ontario through the Ministry of Research and Innovation.

\end{document}